\documentclass[twocolumn,showpacs,%
  aps,superscriptaddress,%
  eqsecnum,prd,notitlepage,showkeys,
  footinbib]{revtex4-1}

\usepackage{amssymb}
\usepackage{amsmath}
\usepackage{graphicx}
\usepackage{dcolumn}
\usepackage{hyperref}
\usepackage{color,units}
\usepackage[dvipsnames]{xcolor} 
\usepackage{aas_macros}
\usepackage{lineno}
\usepackage{xspace}

\usepackage[normalem]{ulem} 

\newcommand{\bilby}{{\sc Bilby}\xspace}
\newcommand{\lal}{{\sc LAL}\xspace}
\newcommand{\lalinference}{{\sc LALInference}\xspace}

\newcommand{\lalsimulation}{{\sc LALSimulation}\xspace}
\newcommand{\python}{{\sc Python}}

\renewcommand{\L}{\mathcal{L}}

\begin{document}

\title{
Bilby: A user-friendly Bayesian inference library for gravitational-wave astronomy
}
\author{Gregory Ashton}
    \email{greg.ashton@monash.edu}
\author{Moritz H\"ubner}
    \email{moritz.huebner@monash.edu}
\author{Paul D. Lasky}
    \email{paul.lasky@monash.edu}
\author{Colm Talbot}
    \email{colm.talbot@monash.edu}
\author{Kendall Ackley}
\affiliation{School of Physics and Astronomy, Monash University, Vic 3800, Australia}
\affiliation{OzGrav: The ARC Centre of Excellence for Gravitational Wave Discovery, Clayton VIC 3800, Australia}

\author{Sylvia Biscoveanu}
\affiliation{LIGO, Massachusetts Institute of Technology, Cambridge, MA 02139, USA}
\affiliation{School of Physics and Astronomy, Monash University, Vic 3800, Australia}
\affiliation{OzGrav: The ARC Centre of Excellence for Gravitational Wave Discovery, Clayton VIC 3800, Australia}

\author{Qi Chu}
\affiliation{OzGrav: The ARC Centre of Excellence for Gravitational Wave Discovery, Crawley WA 6009, Australia}
\affiliation{School of Physics, University of Western Australia, Crawley, Western Australia 6009, Australia}

\author{Atul Divarkala}
\affiliation{Department of Physics, University of Florida, 2001 Museum Road, Gainesville, FL 32611-8440, USA}
\affiliation{School of Physics and Astronomy, Monash University, Vic 3800, Australia}
\affiliation{OzGrav: The ARC Centre of Excellence for Gravitational Wave Discovery, Clayton VIC 3800, Australia}

\author{Paul J. Easter}
\author{Boris Goncharov}
\author{Francisco Hernandez Vivanco}
\affiliation{School of Physics and Astronomy, Monash University, Vic 3800, Australia}
\affiliation{OzGrav: The ARC Centre of Excellence for Gravitational Wave Discovery, Clayton VIC 3800, Australia}

\author{Jan Harms}
\affiliation{Gran Sasso Science Institute (GSSI), I-67100 LAquila, Italy}
\affiliation{INFN, Laboratori Nazionali del Gran Sasso, I-67100 Assergi, Italy}

\author{Marcus E. Lower}
\affiliation{Centre for Astrophysics and Supercomputing, Swinburne University of Technology, Hawthorn VIC 3122, Australia}
\affiliation{Ozgrav, Swinburne University of Technology, Hawthorn VIC 3122, Australia}
\affiliation{School of Physics and Astronomy, Monash University, Vic 3800, Australia}

\author{Grant D. Meadors}
\affiliation{School of Physics and Astronomy, Monash University, Vic 3800, Australia}
\affiliation{OzGrav: The ARC Centre of Excellence for Gravitational Wave Discovery, Clayton VIC 3800, Australia}

\author{Denyz Melchor}
\affiliation{California State University Fullerton, Fullerton, CA 92831, USA}
\affiliation{School of Physics and Astronomy, Monash University, Vic 3800, Australia}
\affiliation{OzGrav: The ARC Centre of Excellence for Gravitational Wave Discovery, Clayton VIC 3800, Australia}

\author{Ethan Payne}
\affiliation{School of Physics and Astronomy, Monash University, Vic 3800, Australia}
\affiliation{OzGrav: The ARC Centre of Excellence for Gravitational Wave Discovery, Clayton VIC 3800, Australia}

\author{Matthew D. Pitkin}
\affiliation{SUPA, School of Physics \& Astronomy, University of Glasgow, Glasgow G12 8QQ, UK}

\author{Jade Powell}
\affiliation{Centre for Astrophysics and Supercomputing, Swinburne University of Technology, Hawthorn VIC 3122, Australia}
\affiliation{Ozgrav, Swinburne University of Technology, Hawthorn VIC 3122, Australia}

\author{Nikhil Sarin}
\author{Rory J. E. Smith}
\author{Eric Thrane}
\affiliation{School of Physics and Astronomy, Monash University, Vic 3800, Australia}
\affiliation{OzGrav: The ARC Centre of Excellence for Gravitational Wave Discovery, Clayton VIC 3800, Australia}

\begin{abstract}
Bayesian parameter estimation is fast becoming the language of gravitational-wave astronomy.
It is the method by which gravitational-wave data is used to infer the sources' astrophysical properties.
We introduce a user-friendly Bayesian inference library for gravitational-wave astronomy, \bilby.  This python code provides expert-level parameter estimation infrastructure with straightforward syntax and tools that facilitate use by beginners. It allows users to perform accurate and reliable gravitational-wave parameter estimation on both real, freely-available data from LIGO/Virgo, and simulated data.
We provide a suite of examples for the analysis of compact binary mergers and other types of signal model including supernovae and the remnants of binary neutron star mergers.
These examples illustrate how to change the signal model, how to implement new likelihood functions, and how to add new detectors.
\bilby{} has additional functionality to do population studies using hierarchical Bayesian modelling.
We provide an example in which we infer the shape of the black hole mass distribution from an ensemble of observations of binary black hole mergers. 
\end{abstract}

\maketitle


\section{Introduction}
Bayesian inference underpins gravitational-wave science.
Following a detection, Bayesian parameter estimation allows one to estimate the properties of a gravitational-wave source, for example, the masses and spins of the components in a binary merger~\cite[e.g.,][]{abbott16_gw150914_pe,abbott16_gw150914_updatedpe,abbott16_01BBH,abbott18_GW170817_NS_parameters,abbott18_GW170817_properties}.
If the detection involves neutron stars, Bayesian parameter estimation is used to study the properties of matter at nuclear densities via the signature of tidal physics imprinted on the gravitational waveform~\cite{abbott17_gw170817_detection,abbott18_GW170817_properties,abbott18_GW170817_NS_parameters}.
The posterior probability distributions of source parameters such as inclination angle can be used, in turn, to make inferences about electromagnetic phenomena such as gamma-ray bursts~\cite[e.g.,][]{abbott17_gw170817_gwgrb}.
Such parameter estimation is also used to measure cosmological parameters such as the Hubble constant~\cite{abbott17_gw170817_Hubble}.
By combining data from multiple detections, Bayesian inference is used to understand the population properties of gravitational-wave sources~\cite[e.g.,][and references therein]{abbott16_01BBH, talbot18, Wysocki2018, smith18, farr18, taylor18, Roulet2018}, which is providing insights into stellar astrophysics.
By extending the gravitational-wave signal model, Bayesian inference is used to test general relativity and look for evidence of new physics~\cite{abbott16_gw150914_testingGR,abbott17_gw170814_detection,abbott17_gw170817_gwgrb,abbott18_cw_polarisations,abbott18_sgwb_polarisations}

The field of gravitational-wave astronomy is growing rapidly.
We have entered the ``open data era,'' in which gravitational-wave data has become publicly available~\cite{gwosc}.
Since Bayesian parameter estimation is central to gravitational-wave science, there is a need for a robust, user-friendly code that can be used by both gravitational-wave novices and experts alike.

The primary tool currently used by the LIGO and Virgo collaborations for parameter estimation of gravitational-wave signals is \lalinference{}~\cite{veitch15}.  This pioneering code enabled the major gravitational-wave discoveries achieved during the first two LIGO observing runs~\cite[e.g.,][]{abbott16_gw150914_pe,abbott16_gw150914_updatedpe,abbott16_01BBH,abbott18_GW170817_NS_parameters,abbott18_GW170817_properties}.  The code itself is now almost a decade old, and years of development have made it hard for beginners to learn, and difficult for experts to modify and adapt to new challenges.   More recently, {\sc PyCBC Inference}~\cite{biwer18} was released; a modern, \python-based toolkit designed for compact binary coalescence parameter estimation.  This package provides access to several different samplers and builds on the {\sc PyCBC} package~\cite{nitz18} -- an open-source toolkit for gravitational-wave astronomy.

We introduce \bilby, a user-friendly parameter-estimation code for gravitational-wave astronomy.
\bilby{} provides expert-level parameter estimation infrastructure with straightforward syntax and tools that facilitate use by beginners.
For example, with minimal user effort, users can download and analyze publicly-available LIGO and Virgo data to obtain posterior distributions for the astrophysical parameters associated with recent detections of binary black holes~\cite{abbott16_gw150914_detection,abbott16_gw151226_detection,abbott16_01BBH,abbott17_gw170104_detection,abbott17_gw170608_detection,abbott17_gw170814_detection} and the binary neutron star merger~\cite{abbott17_gw170817_detection}. 

One key functional difference between \bilby{} and \lalinference/{\sc PyCBC Inference} is its modularity and adaptability. The core library is not specific to gravitational-wave science and has uses outside of the gravitational-wave community.  Ongoing projects include astrophysical inference in multimessenger astronomy, pulsar timing, and x-ray observations of accreting neutron stars.  The gravitational-wave specific library is also built in a modular way, enabling users to easily define their own waveform models, likelihood functions, etc.  This implies \bilby{} can be used for more than studying compact binary coalescences---see Sec.~\ref{sec:signalmodels}.  The modularity further ensures the code will be sufficiently extensible to suit the future needs of the gravitational-wave community. Moreover, we believe the wider astrophysics inference community will find the code useful by virtue of having a common interface and ideas that can be easily adapted to a range of inference problems. 

The remainder of this paper is structured to highlight the versatile, yet user-friendly nature of the code.
To that end, the paper is example driven.
We assume familiarity with the mathematical formalism of Bayesian inference and parameter estimation (priors, likelihoods, evidence, etc.) as well as familiarity with gravitational-wave data analysis (antenna-response functions, power spectral densities, etc.).
Readers looking for an introduction to Bayesian inference in general are referred to Ref.~\cite{Skilling04}, while gravitational-wave specific introductions to inference can be found in Refs.~\cite{veitch15,thrane18}.  Section~\ref{sec:designphilosophy} describes the \bilby{} design philosophy, and Sec.~\ref{sec:codeoverview} provides an overview of the code including installation instructions in Sec.~\ref{sec:installation}.
Subsequent sections show worked examples.
The initial examples are the sort of simple calculations that we expect will be of interest to most casual readers.
Subsequent sections deal with increasingly complex applications that are more likely of interest to specialists.

The worked examples are as follows.
Section~\ref{sec:cbc} is devoted to compact binary coalescences.
In~\ref{sec:gw150914}, we carry out parameter estimation with publicly-available data to analyze GW150914, the first ever gravitational-wave event.
In~\ref{sec:BBHinject}, we study a simulated binary black hole signal added to Monte Carlo noise.
In~\ref{sec:bns}, we study the matter effects encoded in the gravitational waveforms of a binary neutron star inspiral.
In~\ref{waveforms}, we show how it is possible to add more sophisticated gravitational waveform phenomenology, for example, by including memory, eccentricity, and higher order modes.
In~\ref{sec:Australia}, we study an extended gravitational-wave network with a hypothetical new detector.

Section~\ref{sec:signalmodels} is devoted to signal models for sources that are not compact binary coalescences.
In~\ref{sec:SNe}, we perform model selection for gravitational waves from a core collapse supernova.
In~\ref{remnant}, we study the case of a post-merger remnant.
Section~\ref{sec:hyperpe} is devoted to {\em hyper}-parameterization, a technique used to study the population properties of an ensemble of events.
Closing remarks are provided in Section~\ref{conclusions}.

\section{\bilby{} Design Philosophy}\label{sec:designphilosophy}
Three goals guide the design choices of \bilby.
First, we seek to provide a parameter-estimation code that is sufficiently powerful to serve as a workhorse for expert users.
Second, we aim to make the code accessible for novices, lowering the bar to work on gravitational-wave inference.
Third, we desire to produce a code that will age gracefully; advances in gravitational-wave astronomy and Bayesian inference can be incorporated straightforwardly without resort to inelegant workarounds or massive rewrites.
To this end, we adhere to a design philosophy, which we articulate with four principles.
\begin{itemize}
    \item {\bf Modularity.} Wherever possible, we seek to modularize the code and follow the abstraction principle~\cite{pierce02}, reducing the amount of repeated code and easing development.  For example, the sampler is a modularized object, so if a problem is initially analyzed using the {\sc PyMultiNest}~\cite{pymultinest} sampler for example, one can easily switch to the {\sc emcee}~\cite{emcee} sampler or even a custom-built gravitational-wave sampler.  For example, \bilby{} accesses samplers through a common interface; as a result it is trivial to easily switch between samplers to compare performance or check convergence issues. 
    \item {\bf Consistency.} We enforce strict style guidelines, including adherence to the {\sc pep8} style guide for \python{}~\footnote{\url{https://www.python.org/dev/peps/pep-0008/}}. As a result, the code is relatively easy-to-follow and intuitive. In order to maintain integrity of the code while responding to the needs of a large and active user base, we employ {\sc GitLab}'s merge request feature. Updates require approval by two experts. The {\sc pep8} protocol is enforced using continuous integration.
    \item {\bf Generality.} Wherever possible, we keep the code as general as possible. For example, the gravitational-wave package is separate from the package that passes the likelihood and prior to the sampler. This generality provides flexibility. For example, in Section~\ref{sec:hyperpe}, we show how \bilby{} can be used to carry out population inference, even though the likelihood function is completely different to the one used for gravitational-wave parameter estimation. Moreover, a general design facilitates the transfer of ideas into and out of gravitational-wave astronomy from the greater astro-statistics community.
    \item {\bf Usability.} We observe that historically, people find it difficult to get started with gravitational-wave inference. In order to lower the bar, we endeavour to make basic things doable with very few lines of code. We provide a large number of tutorials that can serve as a blueprint for a large variety of real-world problems. Finally, we endeavour to follow the advice of the {\sc pep20} style guide for \python{}~\footnote{\url{https://www.python.org/dev/peps/pep-0020/}}: ``There should be one---and preferably only one---obvious way to do it.'' In other words, once users are familiar with the basic layout of \bilby, they can intuit where to look if they want to, for example, add a new detector (see Section~\ref{sec:Australia}) or include non-standard polarization modes.
\end{itemize}

\section{Code Overview}\label{sec:codeoverview}
\subsection{Installation}\label{sec:installation}
\bilby{} is open-source, MIT licensed, and written in python.
The simplest installation method is through PyPI~\footnote{\url{https://pypi.org/project/BILBY/}}.
The following command installs from the command line:
\begin{verbatim}
    $ pip install bilby
\end{verbatim}
This command downloads and installs the package and dependencies.  The source-code itself can be obtained from the {\tt git} repository~\cite{bilbygit}, which also houses an issue tracker and merge-request tool for those wishing to contribute to code development. Documentation about code installation, functionality, and user syntax is also provided~\cite{bilbydoc}.  Scripts to run all examples presented in this work are provided in the {\tt git} repository.

\subsection{Packages}
\bilby{} has been designed such that logical blocks of code are separated and, wherever possible, code is abstracted away to allow future re-use by other models.  At the top level, \bilby{} has three packages: {\tt core}, {\tt gw}, and {\tt hyper}.
The {\tt core} package contains the key functionalities.
It passes the user-defined priors and likelihood function to a sampler, harvests the posterior samples and evidence calculated by the sampler, and returns a {\tt result} object providing a common interface to the output of any sampler along with information about the inputs.
The {\tt gw} package contains gravitational-wave specific functionality, including waveform models, gravitational-wave specific priors and likelihoods.
The {\tt hyper} package contains functionality for the hierarchical Bayesian inference (see Sec.~\ref{sec:hyperpe}).
A flowchart showing the dependency of different packages and modules is available on the {\tt git} repository~\cite{bilbygit}.

\subsubsection{The {\tt core} package}~\label{sec:core}
The {\tt core} package provides all of the code required for general problems of inference.  It provides a unified interface to several different samplers listed below, standard sets of priors including arbitrary user-defined options, and a universal result object that stores all important information from a given simulation.

Prior and likelihood functions are implemented as classes, with a number of standard types implemented in the {\tt core} package: e.g., the {\tt Normal}, {\tt Uniform}, and {\tt LogUniform} priors, and {\tt GaussianLikelihood}, {\tt PoissonLikelihood}, and {\tt ExponentialLikelihood} likelihoods.  One can write their own custom prior and likelihood functions by writing a new class that inherits from the parent {\tt Prior} or {\tt Likelihood}, respectively.  The user only needs to define how the new prior or likelihood is instantiated and calculated, with all other house-keeping logic being abstracted away from the user.

The prior and likelihood are passed to the function {\sc\verb!run_sampler!}, which allows the user to quickly change the sampler method between any of the pre-wrapped samplers, and to define specific run-time requirements such as the number of live points, number of walkers, etc.  Pre-packaged samplers include Markov Chain Monte Carlo Ensemble samplers {\tt emcee}~\cite{emcee}, {\tt ptemcee}~\cite{ptemcee}, {\tt PyMC3}~\cite{salvatier16}, and Nested samplers~\cite{Skilling04,Skilling06} {\tt MultiNest}~\cite{multinest1,multinest2,multinest3} (through the \python{} implementation {\tt pyMultiNest}~\cite{pymultinest}), {\tt Nestle}~\cite{nestle}, {\tt Dynesty}~\cite{dynesty}, and {\tt CPNest}~\cite{veitch17}.  The {\tt Sampler} class again allows users to specify their own sampler by following the other examples.

Despite the choice of sampler, the output from \bilby{} is universal: an {\tt hdf5} file~\cite{hdf5} that contains all output including posterior samples, likelihood calculations, injected parameters, evidence calculations, etc.  The {\tt Result} object can be used to load in these output files, and also perform common operations such as generating corner plots, and creating plots of the data and maximum posterior fit.

\subsubsection{The {\tt gw} package}
The {\tt gw} package provides the core functionality for parameter estimation specific to transient gravitational waves. Building on the {\tt core} package, this provides prior specifications unique to such problems, e.g., a prior that is uniform in co-moving volume distance, as well as the standard likelihood used when studying gravitational-wave transients~(\cite[e.g., see][]{veitch15} and Eq.~\ref{eq:gaussian}), defined as the {\tt GravitationalWaveTransient} class.  The {\tt gw} package also provides an implementation of current gravitational-wave detectors in the detector module, including their location and orientation, as well as different noise power spectral densities for both current and future instruments.  Standard waveform approximants are also included in the {\tt source} module, which are handled through the \lalsimulation{} package~\cite{LALSuite}.

The {\tt gw} package also contains a set of tools to load, clean and analyse gravitational-wave data.  Many of these functions are built on the {\tt GWpy}~\cite{gwpy} code base, which are contained within {\tt \verb!bilby.gw.detector!} and primarily accessed by instantiating a list of {\tt Interferometer} objects.  This functionality also allows one to implement their own gravitational-wave detector by instantiating a new {\tt Interferometer} object---we show an explicit example of this in Sec.~\ref{sec:Australia}.

\subsection{The {\tt hyper} package}
The {\tt hyper} package contains all required functionality to perform hierarchical Bayesian inference of populations.  This includes both a {\tt Model} module and a {\tt HyperparameterLikelihood} class.  This entire package is discussed in more detail in Sec.~\ref{sec:hyperpe}.

\section{Compact Binary Coalescence}\label{sec:cbc}
In this section, we show a suite of \bilby{} examples analyzing binary black hole and binary neutron star signals.

We employ a standard Gaussian noise likelihood $\cal L$ for strain data $d$ given source parameters $\theta$~\cite{vandersluys08a,vandersluys08b,veitch08}:
\begin{align}\label{eq:gaussian}
    \ln {\cal L}(d|\theta) = -\frac{1}{2}\sum_k 
    \left\{\frac{\left[d_k-\mu_k(\theta)\right]^2}{\sigma_k^2}
    + \ln\left(2\pi\sigma^2_k\right)\right\},
\end{align}
where $k$ is the frequency bin index, $\sigma$ is the noise amplitude spectral density, and $\mu(\theta)$ is the waveform.
The waveform is a function of the source parameters $\theta$, which consist of (at least) eight intrinsic parameters (primary mass $m_1$, secondary mass $m_2$, primary spin vector $\vec{S}_1$, secondary spin vector $\vec{S}_2$) and seven extrinsic parameters (luminosity distance $d_L$, inclination angle $\iota$, polarization angle $\psi$, time of coalescence $t_c$, phase of coalescence $\phi_c$, right ascension and declination ra and dec, respectively.  Table~\ref{tab:bbhpriors} shows the default priors implemented for binary black hole systems.  We show how these priors can be called in Secs.~\ref{sec:gw150914} and~\ref{sec:BBHinject}.
Unless otherwise specified, $\mu(\theta)$ is given using the {\tt IMRPhenomP} approximant~\cite{IMRPhenomP}.
However, the approximant can be easily changed; see Secs.~\ref{sec:BBHinject} and ~\ref{sec:bns}.  Moreover, it is relatively simple to sample in different parameters than those listed above (e.g., chirp mass and mass ratio instead of $m_1$ and $m_2$); examples for doing this are provided in the {\tt git} repository~\cite{bilbygit}.

\begin{table}
\caption{\label{tab:bbhpriors} Default binary black hole priors.  The intrinsic variables are the two black hole masses $m_{1,2}$, their dimensionless spin magnitudes $a_{1,2}$, the tilt angle between their spins and the orbital angular momentum $\theta_{1,2}$, and the two spin vectors describing the azimuthal angle separating the spin vectors $\delta\phi$ and the cone of precession about the system's angular momentum $\phi_{\rm JL}$.  The extrinsic parameters are the luminosity distance $d_L$, the right ascension ra and declination dec, the inclination angle between the observers line of sight and the orbital angular momentum $\iota$, the polarisation angle $\psi$, and the phase at coalescence $\phi_c$.  The phase, spins, and inclination angles are all defined at some reference frequency.  We do not set a default prior for the coalescence time $t_c$.  `sin' and `cos' priors are uniform in cosine and sine, respectively, and `comoving' implies uniform in comoving volume.}
\begin{ruledtabular}
\begin{tabular}{ccccc}
variable & unit & prior & minimum & maximum\\
\hline
$m_{1, 2}$ & M$_\odot$ & uniform & 5 & 100\\
$a_{1, 2}$ & - & uniform & 0 & 0.8\\
$\theta_{1,2}$ & rad. & $\sin$ & 0 & $\pi$\\
$\delta\phi$, $\phi_{\rm JL}$ & rad. & uniform & 0 & $2\pi$\\
$d_L$ & Mpc & comoving & $10^2$& $5\times10^3$\\
ra & rad. & uniform & 0 & $2\pi$\\
dec & rad. & cos & $-\pi/2$ & $\pi/2$\\
$\iota$ & rad. & sin & 0 & $\pi$\\
$\psi$ & rad. & uniform & 0 & $\pi$\\
$\phi_c$ & rad. & uniform & 0 &$2\pi$
\end{tabular}
\end{ruledtabular}
\end{table}

\subsection{GW150914: the onset of gravitational wave astronomy}\label{sec:gw150914}
The first direct detection of gravitational waves occurred on the 14$^{\rm th}$ of September, 2015, when the two LIGO detectors~\cite{LIGO} in Hanford, Washington and Livingston, Louisiana detected the coalescence of a binary black hole system~\cite{abbott16_gw150914_detection}.  The gravitational waves swept through the two detectors with a $6.9^{+0.5}_{-0.4}$ ms time difference which, when combined with polarization information, allowed for a sky-location reconstruction covering an annulus of $590$ deg$^2$~\cite{abbott16_gw150914_detection}.
The initially-published masses of the colliding black holes were given as $36^{+5}_{-4}\,M_\odot$ and $29^{+4}_{-4}\,M_\odot$~\cite{abbott16_gw150914_pe}.
Subsequent analyses with more accurate precessing waveforms constrained the masses to be
$35^{+5}_{-3}\,M_\odot$ and $30^{+3}_{-4}\,M_\odot$ at 90\% confidence~\cite{abbott16_gw150914_updatedpe}.
The distance to the source is determined to be $\unit[440^{+160}_{-180}]{Mpc}$~\cite{abbott16_gw150914_updatedpe}.

In this example, we use \bilby{} to reproduce the parameter estimation results for GW150914.
The data for published LIGO/Virgo events is made available through the Gravitational Wave Open Science Center~\cite{gwosc}.  Built-in \bilby{} functionality downloads and parses this data.  We begin with the following two lines.
\begin{widetext}
\begin{verbatim}
    >>> import bilby
    >>> interferometers = bilby.gw.detector.get_event_data("GW150914")
\end{verbatim}
The first line of code imports the \bilby{} code-base into the \python{} environment.  The second line returns a set of objects that contain the relevant data segments and associated data products relevant for the analysis for both the LIGO Hanford and Livingston detectors.
By default, \bilby{} downloads and windows the data.  A local copy of the data is saved along with diagnostic plots of the gravitational-wave strain amplitude spectral density.

In addition to the data, the two key ingredients for any Bayesian inference calculation are the likelihood and the prior.  Default sets of priors can be called from the {\tt gw.prior} module, and we also employ the default Gaussian noise likelihood (Eq.~\ref{eq:gaussian}). 
\begin{verbatim}
    >>> prior = bilby.gw.prior.BBHPriorDict(filename="GW150914.prior")
    >>> likelihood = bilby.gw.likelihood.get_binary_black_hole_likelihood(interferometers)
\end{verbatim}
The above code calls the GW150914 prior, which differs from the priors described in Tab.~\ref{tab:bbhpriors} in two main ways.  Firstly, to speed up the running of the code it restricts the mass priors to between 30 and 50 M$_\odot$ for the primary mass, and 20 and 40 M$_\odot$ for the secondary mass.  Moreover, this prior call restricts the time of coalescence to 0.1 seconds before and after the known coalescence time.  One can revert to the priors in Tab.~\ref{tab:bbhpriors} by replacing the above file call with {\tt\verb!filename="binary_black_holes.prior"!}, but this would require separately setting a prior for the coalescence time.  We show how this can be done in Sec.~\ref{sec:BBHinject}.

The next step is to call the sampler:
\begin{verbatim}    
    >>> result = bilby.core.sampler.run_sampler(likelihood, prior)
\end{verbatim}
This line performs parameter inference using the sampler default {\sc Dynesty}~\cite{dynesty}, with a default 500 live points.  This number can be increased by passing the {\tt nlive=} keyword argument to {\tt\verb!run_sampler()!}.
The sampler returns a list of posterior samples, the Bayesian evidence, and metadata, which is stored in an {\tt hdf5} file.
One may plot a corner plot showing the posterior distribution for all parameters in the model using the command
\begin{verbatim}
    >>> result.plot_corner()
\end{verbatim}
\end{widetext}

The above example code produces posterior distributions that by eye, agree reasonably well with the parameter uncertainty associated with the published distributions for GW150914.  The shape of the likelihood for the extrinsic parameters presents significant challenges for samplers, due to strong degeneracy's between different sky locations, distances, inclination angles, and polarization angles~\cite[see e.g.,][]{raymond14,farr14c}.  For more accurate results, we use the nested sampling package {\sc CPNest}~\cite{veitch17}, which is invoked by changing the {\tt\verb!run_sampler!} function above to include the additional argument {\tt\verb!sampler='cpnest'!}.  We also change the number of live points by adding {\tt\verb!nlive=5000!} to the same function, and specify a keyword argument {\tt\verb!maxmcmc=5000!}, which is the maximum number of steps the sampler takes before accepting a new sample.  To resolve the issue with the phase at coalescence, we analytically marginalize over this parameter~\cite[][]{farr14} by adding the optional {\tt\verb!phase_marginalization=True!} argument to the instantiation of the likelihood.  \bilby{} has built in analytic marginalization procedures for the time of coalescence~\cite{farr14} and distance~\cite{singer16a,singer16b}, which can both be invoked using {\tt\verb!time_marginalization=True!} and {\tt\verb!distance_marginalization=True!}, respectively.  These decrease the run time of the code by minimizing the dimensionality of the parameter space.  Posterior distributions can still be determined for these parameters by reconstructing them analytically from the full set of posterior samples~\cite[e.g., see][]{thrane18}.

\begin{figure}[t]
	\centering
    \includegraphics[width=1.0\columnwidth]{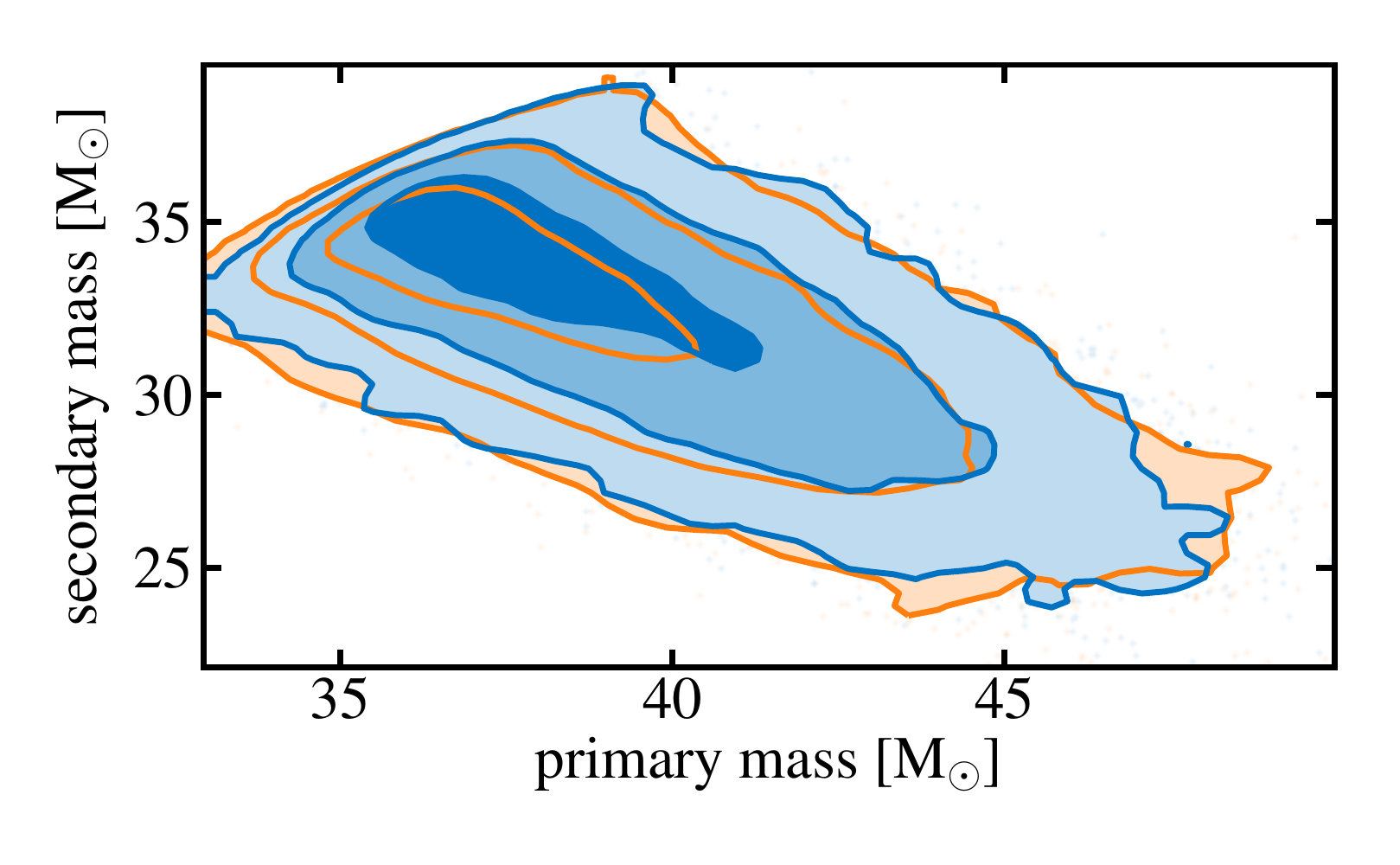}
    \caption{Marginalised posterior source-mass distributions for the first binary black hole merger detected by LIGO, GW150914. We show the posterior distributions recovered using \bilby{} (blue), and those using \lalinference{}~(orange), using open data from the Gravitational Wave Open Science Centre~\cite{gwosc}.  The five lines of \bilby{} code required for reproducing the posteriors are shown in Section~\ref{sec:gw150914}.
    }
    \label{fig:150914Masses}
\end{figure}

Using \bilby{} we can plot marginalized distributions by simply passing the {\tt\verb!plot_corner!} function the optional {\tt\verb!parameters=!\ldots} argument.  In Fig.~\ref{fig:150914Masses} we show the marginalized, two-dimensional posterior distribution for the masses of the two black holes as calculated using the above \bilby{} code (shown in blue).  In orange we show the LIGO posterior distributions from Ref.~\cite{abbott16_01BBH}, calculated using the \lalinference{} software~\cite{veitch15}, and hosted on the Gravitational Wave Open Science Center~\cite{gwosc}.

\begin{figure}[t]
	\centering
    \includegraphics[width=1.0\columnwidth]{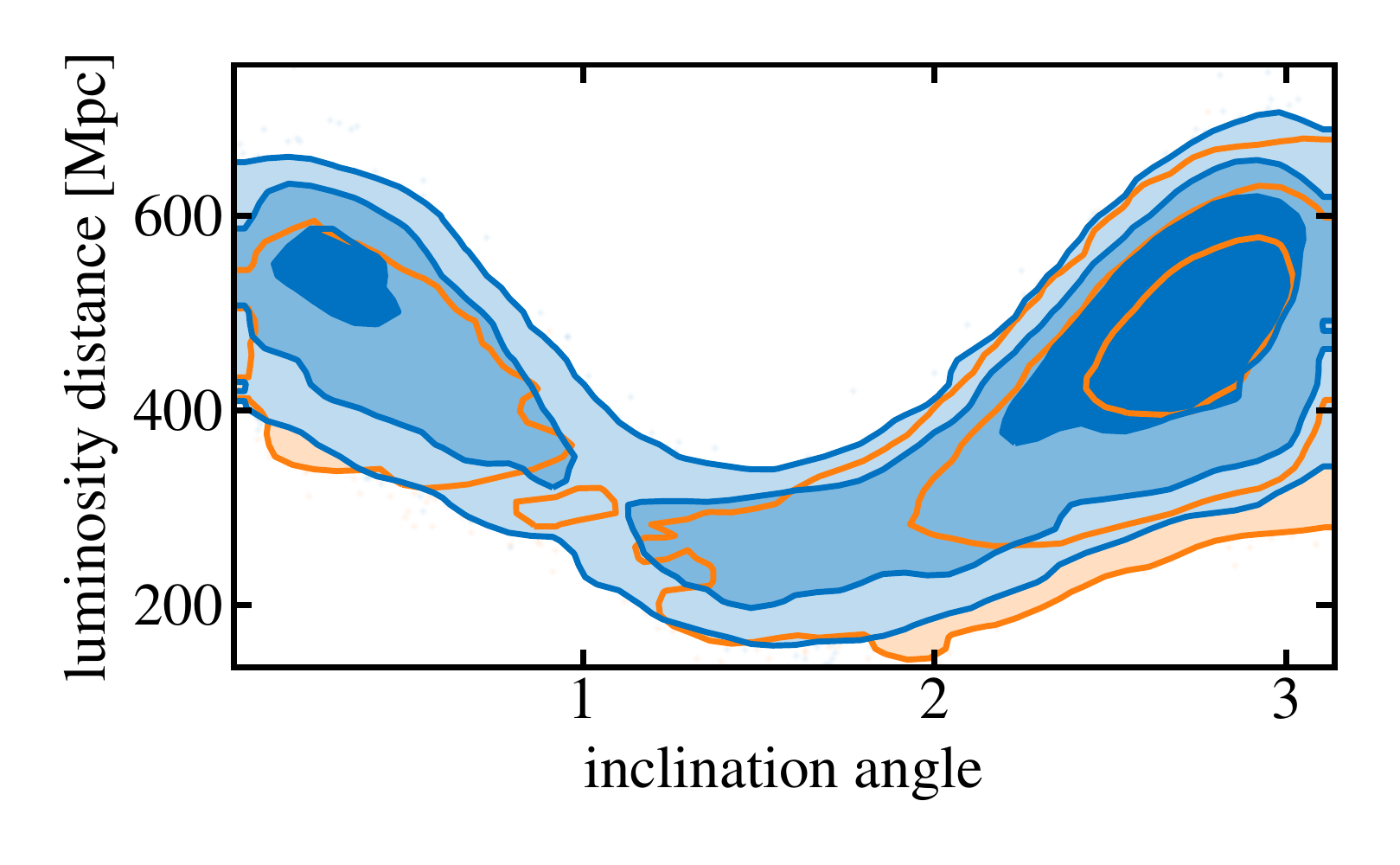}
    \caption{Marginalized posterior distributions on the binary inclination angle and luminosity distance for the first binary black hole merger detected by LIGO, GW150914. We show the posterior distributions recovered using \bilby{} (blue), and those using \lalinference{}~(orange), using open data from the Gravitational Wave Open Science Centre~\cite{gwosc}.
    }
    \label{fig:150914DistInclination}
\end{figure}

\begin{figure}[t]
	\centering
    \includegraphics[width=1.0\columnwidth]{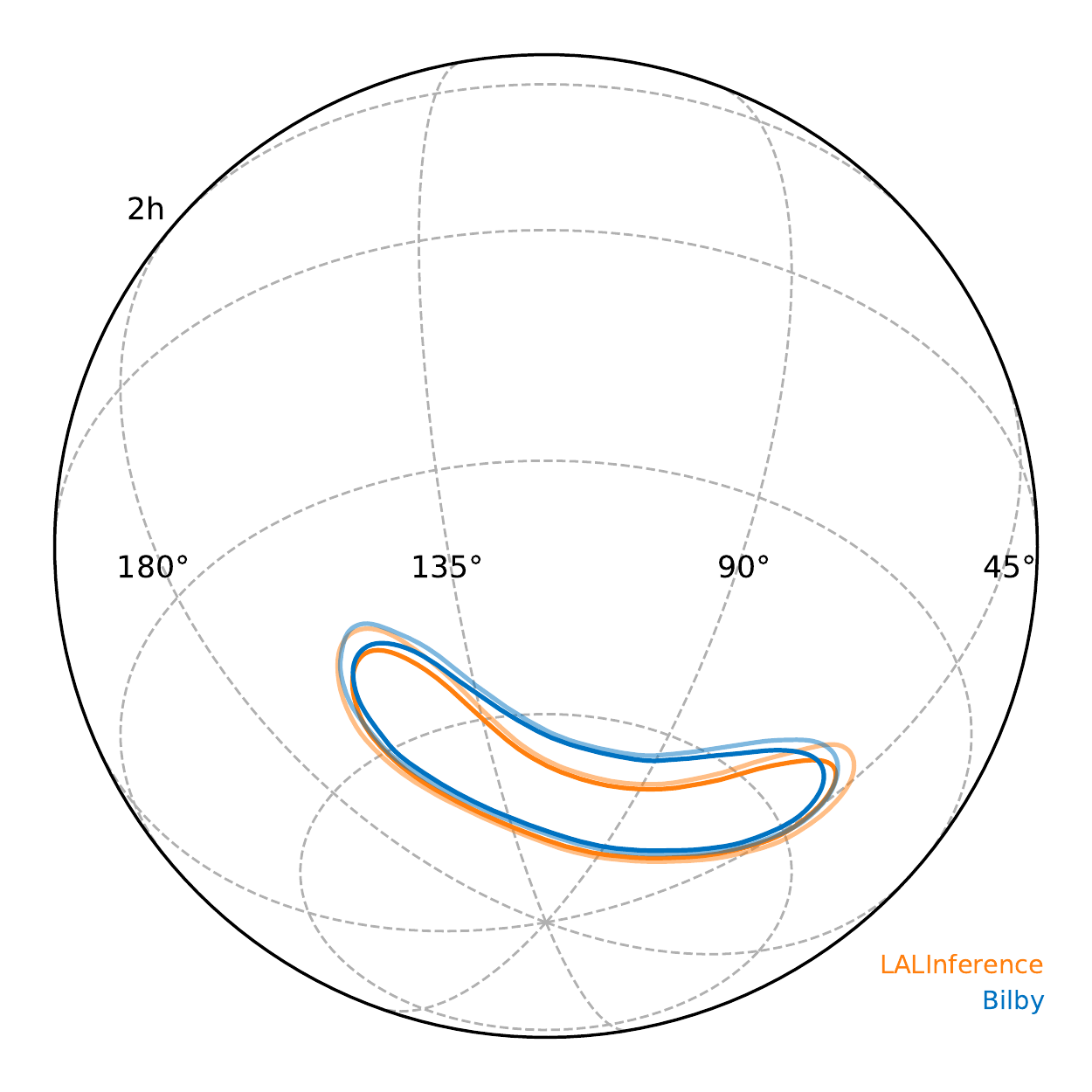}
    \caption{Sky localisation uncertainty for GW150914.  The blue marginalized posterior distributions are those recovered using \bilby, and the orange are those recovered using \lalinference, using open data from the Gravitational Wave Open Science Center~\cite{gwosc}.
    }
    \label{fig:150914skylocalisation}
\end{figure}

In Fig.~\ref{fig:150914DistInclination} we show the marginalized posterior distribution of the luminosity distance and inclination angle, where the \bilby{} posteriors are again shown in blue, and the \lalinference{} posteriors in orange.  Figure~\ref{fig:150914skylocalisation} shows the sky localisation uncertainty for both \bilby{} and \lalinference{}. 

The above example does not make use of detector calibration uncertainty, which is an important feature in LIGO data analysis.  Such calibration uncertainty is built in to \bilby{} using the cubic spline parameterization~\cite{farr14b}, with example usage in the \bilby{} repository.

\subsection{Binary black hole merger injection}\label{sec:BBHinject}
\bilby{} supports both the analysis of real data as in the previous section, as well as the ability to inject simulated signals into Monte Carlo data.  In the following two sections we inject a binary black hole signal and a binary neutron star signal, respectively, showing how one can easily inject and recover signals and their astrophysical properties.

In this first example\footnote{This example is found in the \bilby{} {\tt git} repository at \url{https://git.ligo.org/lscsoft/bilby/blob/master/examples/injection_examples/basic_tutorial.py}.}, we create a binary black hole signal with parameters similar to GW150914~\cite{abbott16_gw150914_pe}, albeit at a luminosity distance of $d_L=2$ Gpc (cf. $d_L\approx400$ Mpc for GW150914). We inject the signal into a network of LIGO-Livingston, LIGO-Hanford~\cite{LIGO} and Virgo interferometers~\cite{virgo}, each operating at design sensitivity.  When doing examples of this nature, it is time intensive to sample over all fifteen parameters in the waveform model.  Therefore, to get quick results that can be run on a laptop, we only sample over four parameters in the waveform model: the two black-hole masses $m_{1,2}$, the luminosity distance $d_L$, and the inclination angle $\iota$.  \bilby{} supports simple functionality to limit or extend the number of parameters included in the likelihood calculation, as shown below.

We begin by setting up a {\tt WaveformGenerator} object using a frequency domain strain model that takes the signal injection parameters and specific waveform arguments such as the waveform approximant as arguments.  The {\tt WaveformGenerator} also takes data duration and sampling frequency as input parameters.  With the source model defined, we now instantiate an {\tt interferometer} object that takes the strain signal from the {\tt WaveformGenerator} and injects it into a noise realisation of the three interferometers.  One could choose to do a zero-noise simulation by simply including the flag {\tt\verb!zero_noise=True!}.

Priors are set up as in the previous open data example, except we call the {\tt\verb!binary_black_holes.prior!} file instead of the specific prior file for GW150914.  Moreover, to hold all but four of the parameters fixed, we set the value of the prior for those other parameters to the injection value.  For example, setting
\begin{verbatim}
    >>> prior['a_1']=0
\end{verbatim}
sets the prior on the dimensionless spin magnitude of the primary black hole to a delta-function at zero.   

\begin{widetext}
In general, we can change the prior for any parameter with one line of code.  For example, to change the prior on the primary mass to be uniform between $m_1=25$ M$_\odot$ and $35$ M$_\odot$, say, one includes
\begin{verbatim}
    >>> prior['mass_1']=bilby.core.prior.Uniform(minimum=25, maximum=35, unit=r'$M_\odot$')
\end{verbatim}
\bilby{} knows about many different types of priors that can all be called in this way.
\end{widetext}
For this example we are also required to define priors on the coalescence time, which we define to be a uniform prior with minimum and maximum one second either side of the injection time.

The likelihood is again set up similarly to the open-data example of Sec.~\ref{sec:gw150914}, although this time we must pass the {\tt interferometer}, {\tt\verb!waveform_generator!}, and {\tt prior}.  Finally, the sampler can be called in the same way as Sec.~\ref{sec:gw150914}; for this example we use the {\tt pyMultiNest} nested sampler~\cite{pymultinest}.

\begin{figure}[t!]
	\centering
    \includegraphics[width=1.0\columnwidth]{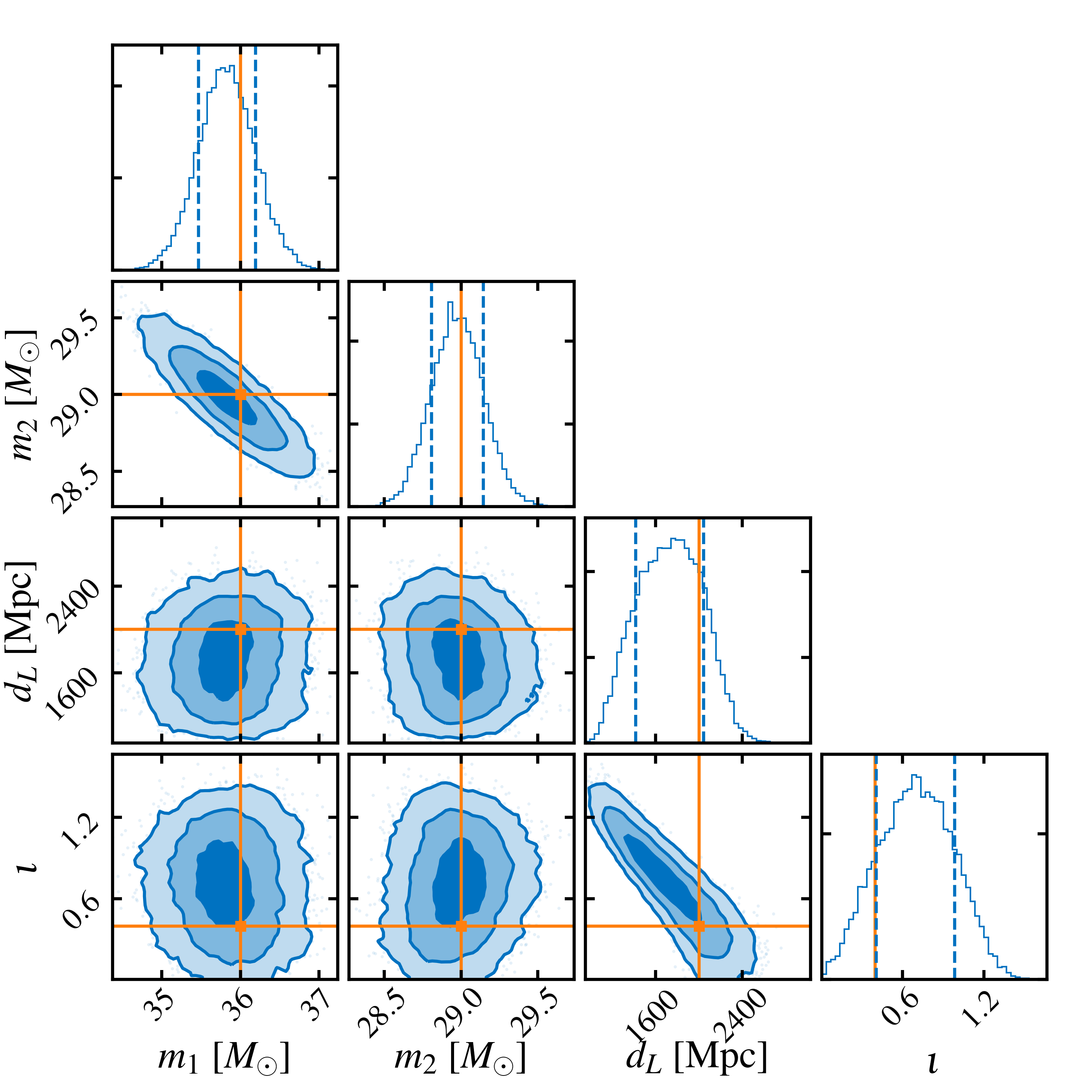}
    \caption{Injecting and recovering a binary black hole gravitational-wave signal with \bilby.  We inject a signal into a three-detector network of LIGO-Livingston, LIGO-Hanford, and Virgo and perform parameter estimation.  The posterior distributions are shown in blue and the injected values in orange.  To speed up the simulation we only search over the two black hole masses $m_1$ and $m_2$, the luminosity distance $d_L$, and the inclination angle $\iota$.}
    \label{fig:basictutorial}
\end{figure}

Figure~\ref{fig:basictutorial} shows the recovered posterior distributions (blue) and the injected parameter values (orange).  For this example, using the {\sc PyMultiNest}~\cite{pymultinest} nested sampling package with 6000 live points took approximately 30 minutes on a laptop to sample fully the four-dimensional parameter space.  The parameters in Fig.~\ref{fig:basictutorial} are recovered well with the usual degeneracy present between the luminosity distance and inclination angle of the source, $d_L$ and $\iota$, respectively.

\subsection{Measuring tidal effects in binary neutron star coalescences}\label{sec:bns}
The first detection of binary neutron star coalescence GW170817 was a landmark event signalling the beginning of multimessenger gravitational-wave astronomy~\cite{abbott17_gw170817_detection, abbott17_gw170817_multimessenger}.  Gravitational-wave parameter estimation of the inspiral is what ultimately determined that both objects were likely neutron stars, and provides the best-yet constraints on the nuclear equation of state of matter at supranuclear densities~\cite{abbott17_gw170817_detection, abbott17_gw170817_multimessenger, abbott17_gw170817_gwgrb}.

One of the key measurements in determining the equation of state from binary neutron star coalescences is that of the tidal parameters.  The dimensionless tidal deformability 
\begin{align}
    \Lambda=\frac{2k_2}{3}\left(\frac{c^2R}{Gm}\right)^5,\label{eq:lambda}
\end{align}
is a fixed parameter for a given equation of state and neutron star mass.  Here, $k_2$ is the second Love number, $R$ and $m$ are the neutron star radius and mass, respectively.  The binary neutron star merger GW170817 provided constraints of $\Lambda_{1.4}=190^{+390}_{-120}$~\cite{abbott18_GW170817_NS_parameters,de18}, where the subscript denotes this is the estimate on $\Lambda$ assuming a 1.4 M$_\odot$ neutron star, and the uncertainty is the 90\% credible interval.

\bilby{} can be used to study neutron star coalescences in both real and simulated data.
We inject a binary neutron star signal using the {\tt TaylorF2} waveform approximant into a three-detector network of the two LIGO detectors and Virgo, all operating at design sensitivity~\footnote{This example is found in the \bilby{} {\tt git} repository at \url{https://git.ligo.org/lscsoft/bilby/blob/master/examples/injection_examples/binary_neutron_star_example.py}.}.  Our injected signal is an $m_{1}=1.3\,{\rm M}_\odot$, $m_{2}=1.5\,{\rm M}_\odot$ binary at $d_L=50$ Mpc with dimensionless spin parameters $a_{1,2}=0.02$, and tidal deformabilities $\Lambda_{1,2}=400$.  Setting up such a system in \bilby{} is equivalent to doing the binary black hole injection study of Sec.~\ref{sec:BBHinject}, except we call the {\tt\verb!lal_binary_neutron_star!} source function, which requires the additional $\Lambda_{1,2}$ arguments.
We also have specific binary neutron star priors; the default set can be called using 
\begin{verbatim}
    >>> priors = bilby.gw.prior.BNSPriorDict()
\end{verbatim}
The standard set of binary neutron star priors are shown in Tab.~\ref{tab:bnspriors}.  In this example we use the {\tt Dynesty} sampler~\cite{dynesty}.

\begin{table}
\caption{\label{tab:bnspriors} Default binary neutron star priors.  $\Lambda_{1,2}$ are the tidal deformability parameters of the primary and secondary neutron star defined in Eq.~\ref{eq:lambda}. For other variable definitions, see Tab.~\ref{tab:bbhpriors}.  Note our commonly-used waveform approximant does not allow misaligned neutron star spins, implying we do not require priors on those spin parameters.}
\begin{ruledtabular}
\begin{tabular}{ccccc}
variable & unit & prior & minimum & maximum\\
\hline
$m_{1, 2}$ & M$_\odot$ & uniform & 1 & 2\\
$a_{1, 2}$ & - & uniform & -0.05 & 0.05\\
$\Lambda_{1,2}$ & - & uniform & 0 & 3000\\
$d_L$ & Mpc & comoving & $10$& $500$\\
ra & rad. & uniform & 0 & $2\pi$\\
dec & rad. & cos & $-\pi/2$ & $\pi/2$\\
$\iota$ & rad. & sin & 0 & $\pi$\\
$\psi$ & rad. & uniform & 0 & $\pi$\\
$\phi_c$ & rad. & uniform & 0 &$2\pi$
\end{tabular}
\end{ruledtabular}
\end{table}

\begin{figure}[t!]
	\centering
    \includegraphics[width=1.0\columnwidth]{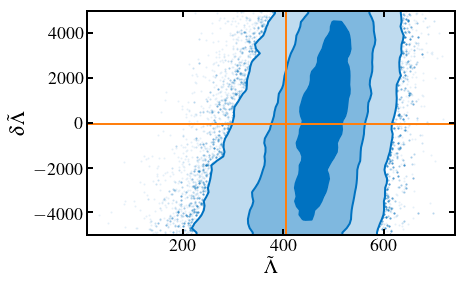}
    \caption{Injecting and recovering a binary neutron star gravitational-wave signal with \bilby.  We inject a signal into the three-detector network, and show here only the marginalized two-dimensional posterior on the two tidal deformability parameters (blue) with the injected values shown in orange.}
    \label{fig:bns}
\end{figure}

The tidal deformability parameters $\Lambda_1$ and $\Lambda_2$ are known to be highly correlated.  The terms that appear explicitly due to the tidal corrections in the phase evolution are instead $\tilde{\Lambda}$ and $\delta\tilde{\Lambda}$~\cite{flanagan08} (for definitions of these parameters, see Eqs. (14) and (15) of Ref.~\cite[][]{lackey15}).  We therefore sample in $\tilde{\Lambda}$ and $\delta\tilde{\Lambda}$, instead of $\Lambda_{1}$ and $\Lambda_{2}$.  Although we sample in all binary neutron star parameters, we show only the two-dimensional marginalized posterior distribution for $\tilde{\Lambda}$ and $\delta\tilde{\Lambda}$ in Fig.~\ref{fig:bns}.   The corresponding injected values of $\tilde{\Lambda}$ and $\delta\tilde{\Lambda}$ are shown as the orange vertical and horizontal lines, respectively.

\subsection{Implementing New Waveforms}\label{waveforms}
The preceding subsections have only given a flavour of what can be achieved with \bilby{} for compact binary coalescences.
It is trivial to implement more complex signal models that include, for example, higher order modes, eccentricity, gravitational-wave memory, non-standard polarizations.
Examples showing different signal models are included in the {\tt git} repository~\cite{bilbygit}.
\bilby{} has already been used in one such application: testing how well the orbital eccentricity of binary black hole systems can be measured with Advanced LIGO and Advanced Virgo~\cite{lower18}.
An example script reproducing those results can be found in the {\tt git} repository~\cite{bilbygit}.  

If a signal model exists in the \lal software~\cite{LALSuite}, then calling that signal model and defining which parameters to include in the sampler is as simple as the above examples.  In Sec.~\ref{sec:signalmodels} we also show how to include a user-defined source model.
Moreover, one is free to define and sample models in either the time or frequency domain.
We include examples for both cases in the {\tt git} repository.
The latter case of using a time-domain source model requires doing little more than selecting the argument {\tt\verb!time_domain_source_model!} in the {\tt WaveformGenerator}, rather than selecting {\tt\verb!frequency_domain_source_model!}.

Of course, one may also want to set up the injection and the sampler using two different waveform models, for example to inject a numerical relativity signal into Monte Carlo data and recover it with a waveform approximant (see also Sec.~\ref{sec:SNe}).  This is possible by simply instantiating two {\tt WaveformGenerator}s, injecting with one and passing the other to the {\tt likelihood}.

\subsection{Adding detectors to the network}\label{sec:Australia}
The full network of ground-based gravitational-wave interferometers will soon consist of the two LIGO detectors in the US, Virgo, LIGO-India~\cite{LIGOIndia} and the KAGRA detector in Japan~\cite{kagra}, all of which are implemented in \bilby. 
A gravitational-wave interferometer is specified by its geographic coordinates, orientation, and noise power spectral density.
By default, \bilby{} includes descriptions of current detectors including LIGO, Virgo, and KAGRA, as well as proposed future detectors, A+~\cite{Aplus}, Cosmic Explorer~\cite{CosmicExplorer}, and the Einstein Telescope~\cite{EinsteinTelescope}.
It is also possible to define new detectors, which is useful for developing the science case for proposals and to optimize the design and placement of new detectors.
Among other things, this can be used in developing the science-case for interferometer design and placement.

\bilby{} provides a common interface to define detectors by their geometry, location, and frequency response.  By way of example, we place a new four-kilometer-arm interferometer in the Shire of Gingin, located outside of Perth, Australia; the current location of the Australian International Gravitational Observatory (AIGO).  We assume a futuristic network configuration of the Australian Observatory together with the two LIGO detectors in Hanford and Livingston, all operating at A+ sensitivity~\cite{Aplus}.  We generate A+ power spectral densities in the same script used to run \bilby{} by using the {\sc pygwinc} software~\cite{pygwinc}, which creates an array containing the frequency and noise power spectral density\footnote{This example is found in the \bilby{} {\tt git} repository at \url{https://git.ligo.org/lscsoft/bilby/blob/master/examples/injection_examples/Australian_detector.py}.} (one could equally use more sophisticated software such as {\sc Finesse}~\cite{finesse} to create more detailed interferometer sensitivity curves).  We then create a new {\tt Interferometer} object using {\tt\verb!bilby.gw.detector.Interferometer()!}, which takes numerous arguments including the position and orientation of the detector, minimum and maximum frequencies, and the power or amplitude noise spectral density.
The noise spectral density can be passed as an ascii file containing the frequency and spectral noise density.
With the new detector defined, one can again calculate a noise realisation and signal injection in a manner similar to what is done in Sec.~\ref{sec:cbc}.

\begin{figure}[t!]
	\centering
    \includegraphics[width=1.0\columnwidth]{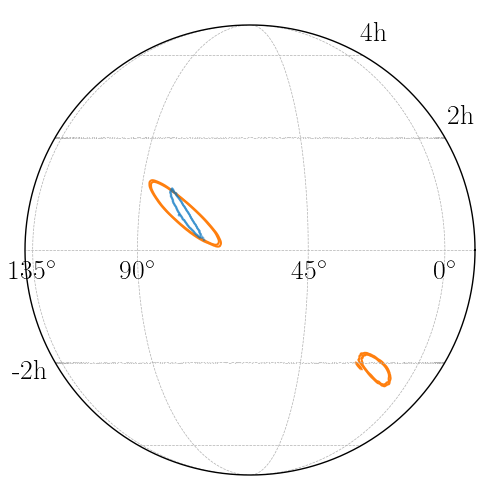}
    \caption{Sky location uncertainty when including a gravitational-wave detector in Gingin, Australia.  Shown are the sky localisations (marginalized two-dimensional posterior distributions) for an injected binary black hole signal using a two-detector network of gravitational-wave interferometers Hanford and Livingston (orange) and a three-detector network that also includes the Australian detector (blue).
    \label{fig:AusIFO}}
\end{figure}

In this example we inject a GW150914-like binary black hole inspiral signal at a luminosity distance of $d_L$=4 Gpc, and recover the masses, sky location, luminosity distance and inclination angle of the system.  In this example we use the {\tt Nestle} sampler~\cite{nestle}.  Figure~\ref{fig:AusIFO} shows the two-dimensional marginalized posterior for the sky-location uncertainty when including (blue) and not including (orange) the Australian detector in Gingin. 
In this instance, the sky localisation uncertainty decreases by approximately a factor four when including the third detector.

While this example includes three detectors, it is straightforward to extend this analysis to an arbitrary detector network.
The likelihood evaluation simply loops over the number of detectors passed to it and multiplies the likelihood for each detector to get a combined likelihood for each point in the parameter space.

\section{Alternative signal models}\label{sec:signalmodels}
Section~\ref{sec:cbc} focuses on compact binary coalescences.  However, the \bilby{} {\tt gw} package enables parameter estimation for any type of signal for which a signal model can be defined.  In this section, we show two illustrative examples: the injection and recovery of a core-collapse supernovae signal, and a much-simplified model of a hypermassive neutron star following a binary neutron star merger.  The former example highlights two key pieces of infrastructure; the ability to inject numerical relativity signals, and to develop ones own source model that is not built into \bilby.  The latter example highlights the use of a different likelihood function that only uses the amplitude of the signal, and throws away the phase information.

\subsection{Supernovae}\label{sec:SNe}
Gravitational-wave signals from core-collapse supernovae are complicated and not well understood in terms of their specific phase evolution.  Numerous techniques have been developed to deal with both detection and parameter estimation.  One such method for the latter problem involves principal component analysis~\cite{logue12, powell16, powell17}, where the signal is reconstructed using a weighted sum of orthonormal basis vectors.  In this example, we inject a gravitational-wave signal from a numerical relativity simulation~\cite{mueller12} and recover the principal components using \bilby\footnote{This example is found in the \bilby{} {\tt git} repository at \url{https://git.ligo.org/lscsoft/bilby/blob/master/examples/supernova_example/supernova_example.py}.}.  

\begin{figure}[t!]
	\centering
    \includegraphics[width=1.0\columnwidth]{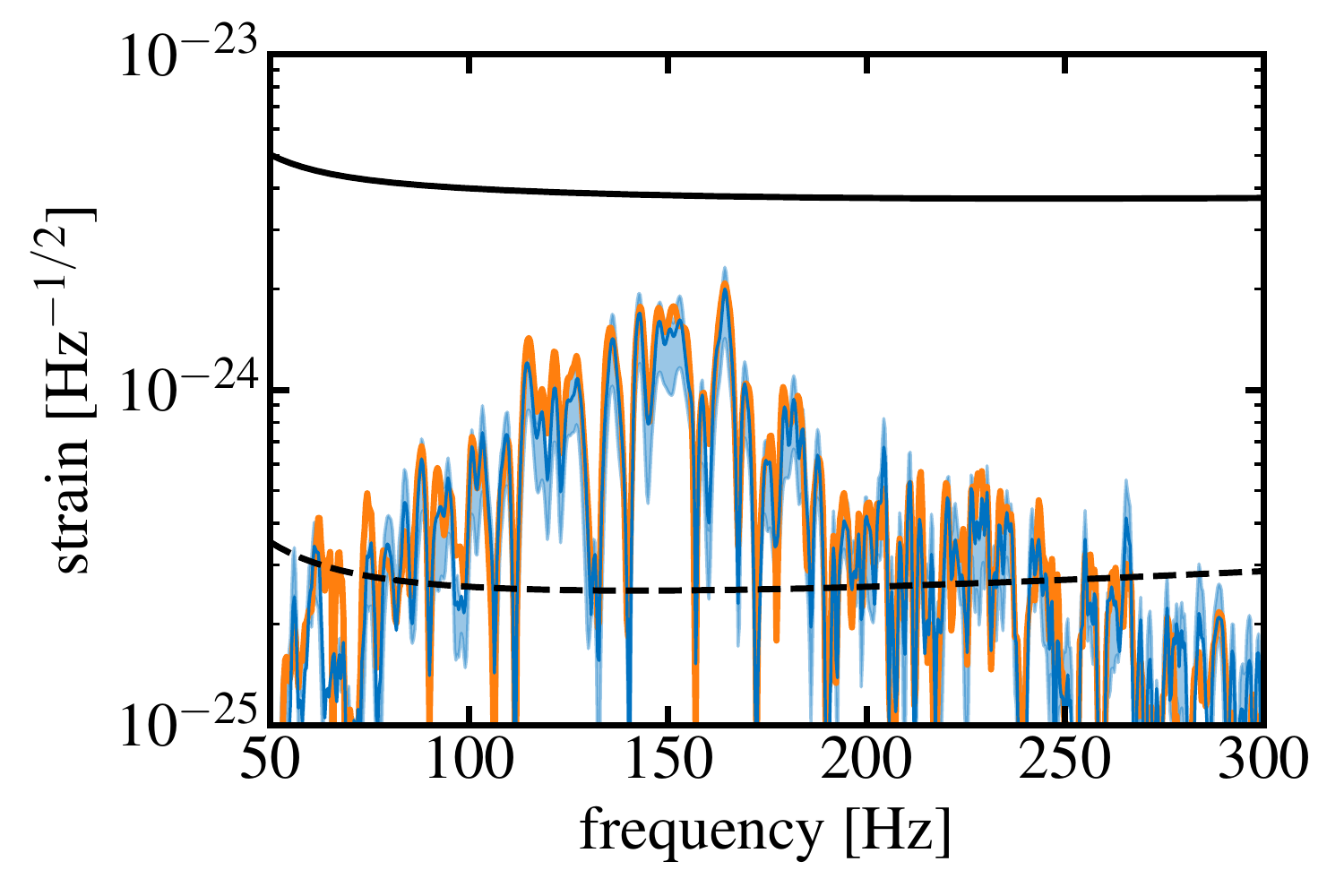}
    \caption{Parameter estimation reconstruction of a numerical relativity supernova signal.  A numerical relativity supernovae signal (orange) is injected into a three-detector network of the two Advanced LIGO detectors and Advanced Virgo, all operating at design sensitivity.  The maximum likelihood reconstruction of the signal is shown in dark blue, and the blue light band shows the superposition of many reconstructed waveforms from the posterior samples.}
    \label{fig:SN}
\end{figure}

The injection is performed by defining a new signal class that, in this case, simply reads in an ascii text file containing the gravitational-wave strain time series.  The injection is then performed in a way akin to the binary black hole and binary neutron star examples in Sec.~\ref{sec:cbc}.  We inject signal L15 from Ref.~\cite{mueller12}, which comes from a three-dimensional simulation of a non-rotating core-collapse supernova with a 15 M$_\odot$ progenitor star.  The signal is injected at a distance of 5 kpc in the direction of the galactic center.  The amplitude spectral density of the injected signal is shown in Fig.~\ref{fig:SN} as the orange trace.

The signal is reconstructed using principal component analysis, such that the strain is expressed as 
\begin{align}
    \tilde{h}(f)=A\sum_{j=1}^k\beta_jU_j(f),\label{eq:PCA}
\end{align}
where $A$ is an amplitude factor, $\beta_j$ and $U_j$ are the complex principal component amplitudes and vectors, respectively.  Equation~(\ref{eq:PCA}) is implemented into \bilby{} as another new signal model that takes the $\beta_j$ coefficients, luminosity distance (which is a proxy for $A$), and sky location as inputs.  Priors for each of the new parameters are established in the same way as the example with the mass in Sec.~\ref{sec:BBHinject}.  In this case, we set $k=5$ and use uniform priors between -1 and 1 for each of the $\beta_j$'s.

Figure~\ref{fig:SN} shows the injected (orange) and recovered (blue) gravitational-wave signal in the frequency domain.  The dark blue curve shows the maximum likelihood curve, and the shaded blue region is a superposition of many reconstructed waveforms from the posterior samples.

\subsection{Neutron star post-merger remnant}\label{remnant}
There are a number of physical scenarios that can occur following the merger of two neutron stars, including the existence of short- or long-lived neutron star remnants.  In the early phases post-merger ($\lesssim$1\,s), these neutron stars are highly dynamic, and can emit significant gravitational radiation potentially observable by Advanced LIGO and Virgo at design sensitivity out to $\sim50$ Mpc~\cite[e.g.,][and references therein]{clark14}.  While the ultimate fate of binary merger GW170817 is unknown, no gravitational waves from a post-merger remnant were found~\cite{abbott17_gw170817_postmerger,abbott18_GW170817_properties}, which is not surprising given the interferometers were not operating at design sensitivity and the distances involved.  

Providing the sensitivity of gravitational-wave interferometers continues to increase,  it is possible a gravitational-wave signal from a post-merger remnant could be detected in the relative near future.  Such a detection would provide an excellent opportunity to understand the nuclear equation of state of matter at extreme densities, as well as the rich physics of these exotic objects~\cite[e.g.,][]{shibata06,baiotti08,read13}.  Parameter inference of such short-lived signals is in its infancy~\cite[e.g., see][]{chatziioannou17}, largely due to the paucity of reliable waveforms~\cite{clark16,easter18}.  This is an ongoing challenge due to the expensive nature of numerical relativity simulations and the complex physics that must be included in such simulations.

Simple models that provide approximate gravitational-wave signals fit to a handful of numerical relativity waveforms exist~\cite{messenger14,bose18,easter18}, which may eventually be used for full parameter inference.  The phase evolution of such numerical relativity simulations is rapid, and very difficult to model~\cite{messenger14,easter18}.  However, it is the frequency content of the signal that carries information about the equation of state and the physics of the remnant~\cite[e.g.,][and references therein]{takami15}.  It is therefore possible that parameter-estimation algorithms may require one to throw away information about the phase, and only keep amplitude spectral content.  Such a process requires a different likelihood function than the one that has been used to this point.
This therefore provides good motivation for showing how to include a different likelihood function in \bilby{} code.

We implement a power-spectral density (``burst'') likelihood
\begin{align}
    \ln\L(|d|\,|\,\theta)=\sum_{i=1}^N & \left[   \ln I_0\left(\frac{|\tilde{h}_i(\theta)||\tilde{d_i}|}{S_n(f_i)}\right)-  \frac{|\tilde{h}_i|^2+|\tilde{d}_i|^2}{2S_n(f_i)}\right.\nonumber\\
     &  +\ln{|\tilde{h}_i(\theta)|}-\ln S_n(f_i)\left. \vphantom{\sum_{i=1}^N    \ln I_0\left(\frac{|a||b|}{c}\right)}\right],\label{eq:PSDLikelihood}
\end{align}
where $I_0$ is the zeroth-order modified Bessel function of the first kind. This requires setting up a new Likelihood class, that contains a {\tt\verb!log_likelihood!} function that reads in the frequency array, noise spectral density and waveform model, and outputs a single likelihood evaluation.
Having defined a new likelihood function, one calls the remaining functions in the usual way; the likelihood function is instantiated and passed to the {\tt\verb!run_sampler()!} command.

\begin{figure}[t!]
	\centering
    \includegraphics[width=1.0\columnwidth]{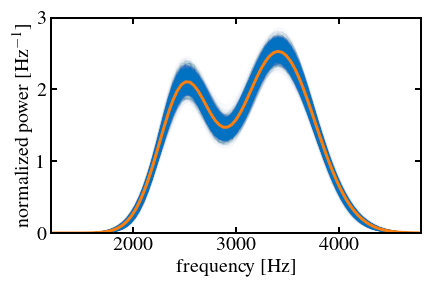}
    \caption{A proxy post-merger gravitational-wave signal from a short-lived neutron star showing the implementation of a different likelihood function in \bilby.  The orange curve is an injected, double-peaked Gaussian signal injected into a constant noise realisation.  The blue band shows the waveform reconstructions from the posterior samples using a power-spectrum likelihood function; i.e., one that only uses the amplitude of the signal and ignores the phase.}
    \label{fig:hmns}
\end{figure}

We inject a double-peaked Gaussian, shown in Fig.~\ref{fig:hmns} as the solid orange curve.  We recover this signal using the same model (with a constant noise spectral density), where we use uniform priors for the amplitudes, widths and frequencies of each of the peaks.  Figure~\ref{fig:hmns} shows the waveform reconstruction for each of the posterior samples, which can be seen to cover the injected signal.  

\section{Population Inference: hyperparameterizations}\label{sec:hyperpe}
Individual detections of binary coalescences can provide stunning insights into various physical and astrophysical questions.   Increased detector sensitivities imply significantly more events will be detected, enabling statements to also be made about ensemble properties of populations~\cite[e.g.,][and references therein]{abbott16_01BBH, talbot18, Wysocki2018, smith18, farr18, taylor18, Roulet2018}.  Extracting information from a population of events is performed using hierarchical Bayesian inference where the population is described by a set of hyper-parameters, $\Lambda$.  \bilby{} has built-in support for calculating $\Lambda$ from multiple sets of posterior samples from individual events.

\bilby{} implements the conventional method whereby the posterior samples ${\theta^{j}_{i}}$ for each event $j$ are re-weighted according to the ratio of the population model prior $\pi(\theta | \Lambda)$ and the sampling prior $\pi(\theta)$ to obtain the hyper-parameter likelihood
\begin{equation}
    \mathcal{L}(h | \Lambda) = \prod^{N}_{j} \frac{\mathcal{Z}_{j}}{n_{j}} \sum^{n_{j}}_{i} \frac{\pi(\theta^{j}_{i} | \Lambda)}{\pi(\theta^{j}_{i})} .
\label{eq:hyperpe}
\end{equation}
Here, $\mathcal{Z}_{j}$ is the Bayesian evidence for the data given the original model and $n_{j}$ is the number of posterior samples in the $j^{\rm th}$ event.

The \bilby{} implementation requires the user to define $\pi(\theta|\Lambda)$ and $\pi(\theta)$ which, along with the set of posterior samples $\theta_i^j$, are passed to the {\tt\verb!HyperparameterLikelihood!} in \bilby's {\tt hyper} package.  The hyperparameter priors are then set up in the usual way, and passed to the standard {\tt\verb!run_sampler!} function.

\begin{figure}
    \centering
    \includegraphics[width=1.0\columnwidth]{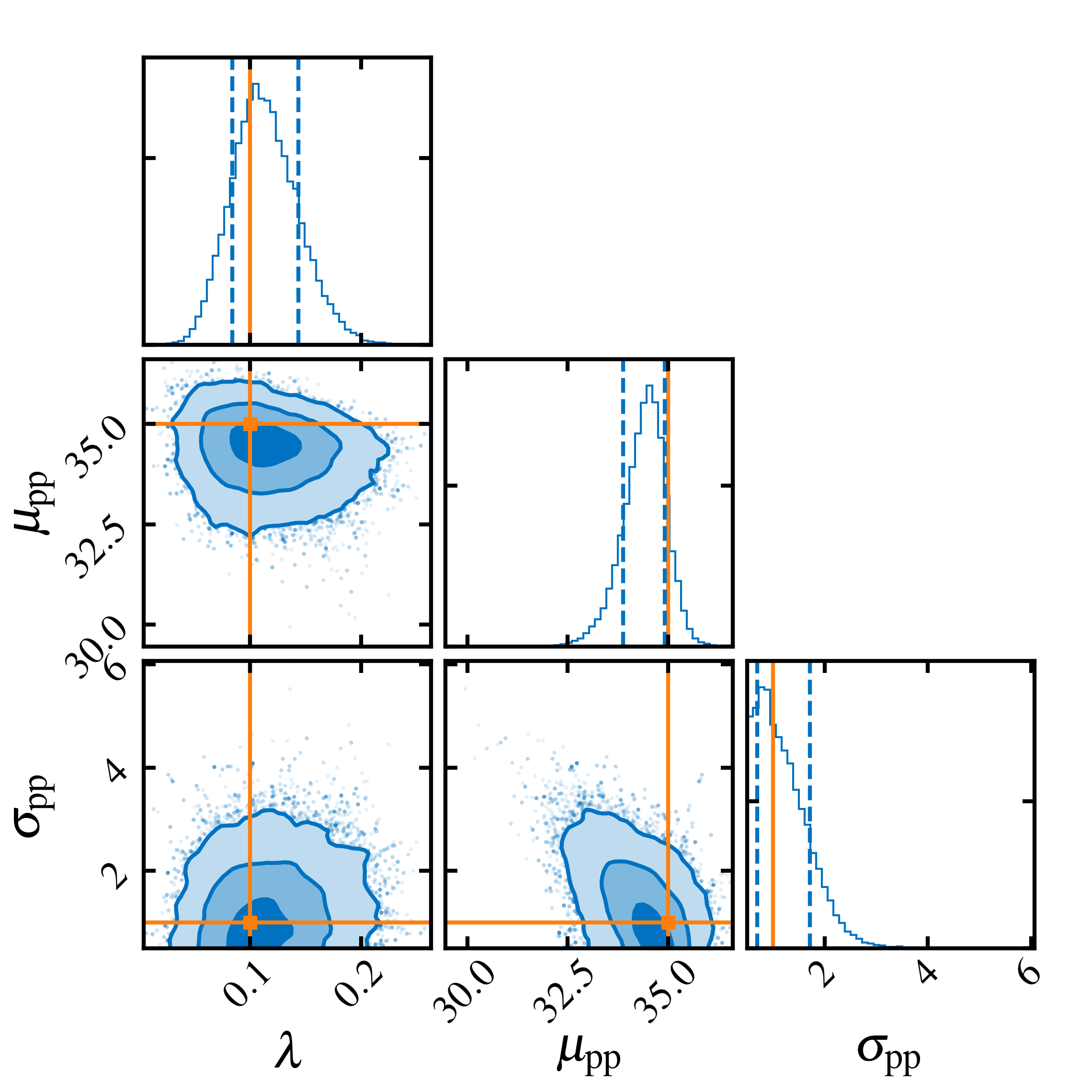}
    \caption{Population modelling with \bilby{} hierarchical Bayesian inference module.  We show the recovery of parameters describing part of the mass distribution of binary black holes using the model described in Ref.~\cite{talbot18}.  The population parameters are drawn from values shown in orange, and the posterior distributions for the hyperparameters shown in blue.  Here, $\lambda$ is the fraction of binaries where the more massive black hole formed through pulsational pair-instability supernovae, $\mu_{\rm pp}$ and $\sigma_{\rm pp}$ are the typical mass of these black holes and the width of the ``PPSN graveyard'', respectively.
    }
    \label{fig:hyper_mass}
\end{figure}

As a demonstration~\footnote{This example is found in the \bilby{} {\tt git} repository at \url{https://git.ligo.org/lscsoft/bilby/blob/master/examples/other_examples/hyper_parameter_example.py}.} of this method we reproduce results \cite{talbot18} recovering parameters describing a postulated excess of black holes due to pulsational pair-instability supernovae (PPSN)~\cite{Heger2003,Woosley2015}.
The posterior distribution for the hyperparameters determining the abundance and characteristic mass of black holes formed through this mechanism are shown in Fig.~\ref{fig:hyper_mass}.
The hyperparameter $\lambda$ is the fraction of binaries where the more massive black hole formed through PPSN, $\mu_{\rm pp}$ is the typical mass of these black holes and $\sigma_{\rm pp}$ determines the width of the ``PPSN graveyard''.

This model contains seven additional hyperparameters describing the remainder of the distribution of black hole masses that we hold fixed for the purposes of this example.
Additional hyperparameters may be added straightforwardly.

\section{Analysis of arbitrary data: an example}\label{sec:logo}
\bilby{} is more than a tool for gravitational-wave astronomy; it can also be used as a generic and versatile inference package.
In the documentation examples, we demonstrate how \bilby{} can be applied to generic time-domain data from radioactive decay processes.
Furthermore, \bilby{} is currently being used to analyse radio and x-ray data from neutron stars, and to study multi-messenger signals associated with binary neutron star mergers.
Here we show an example that calculates posterior distributions for one of the letters in the \bilby{} logo.

\begin{figure}[t!]
	\centering
    \includegraphics[width=1.0\columnwidth]{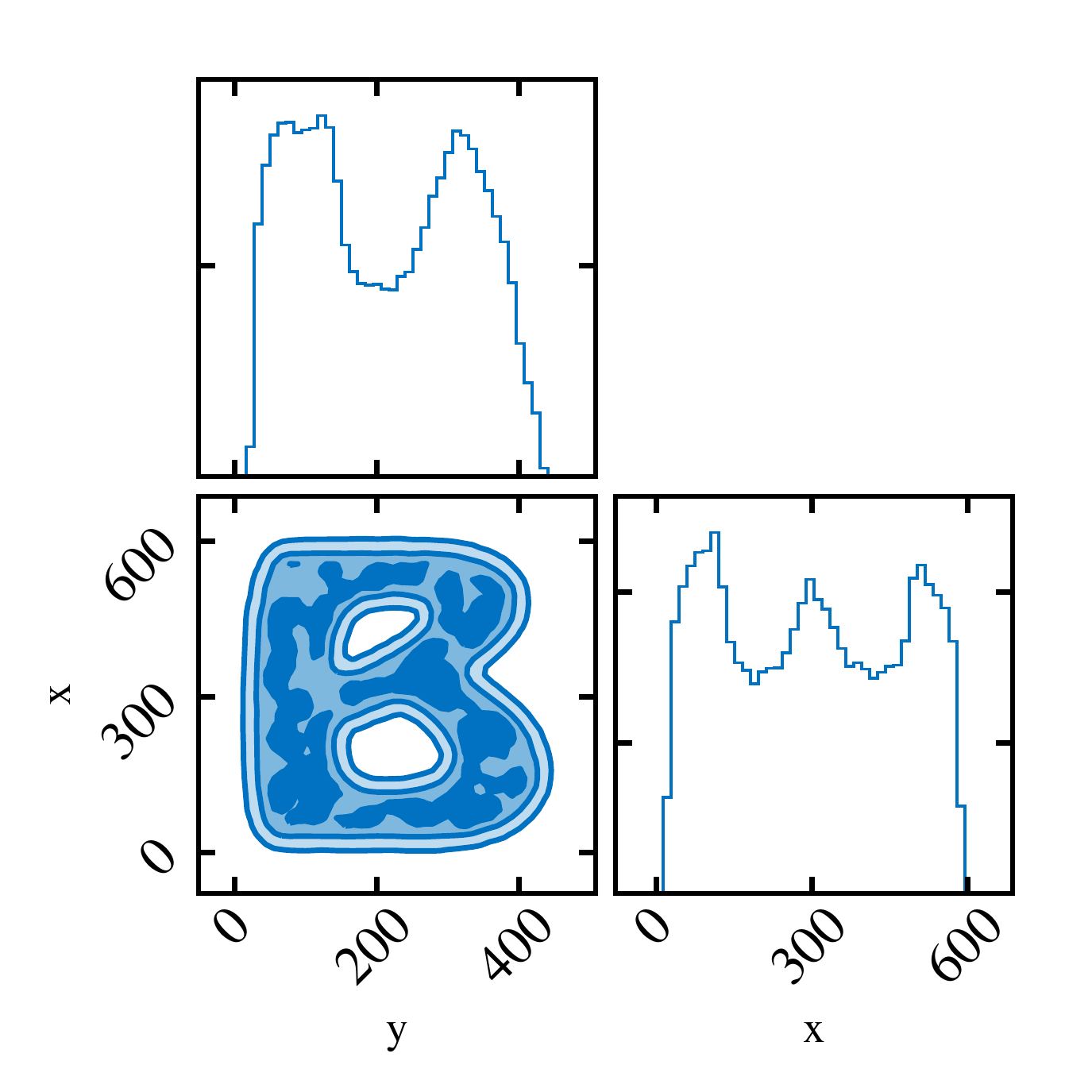}
    \caption{The `B' from the \bilby{} logo, generated using the \bilby{} package; see Sec.~\ref{sec:logo}}
    \label{fig:logo}
\end{figure}

\begin{figure}[t!]
    \centering
    \includegraphics[width=1.0\columnwidth]{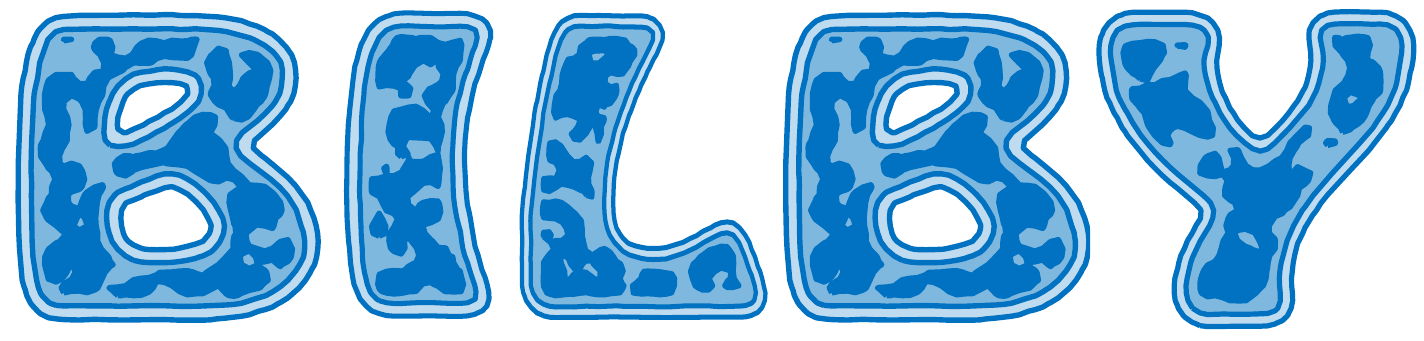}
    \caption{All letters from the \bilby{} logo, generated using the \bilby{} package; see Sec.~\ref{sec:logo}}
    \label{fig:logo_word}
\end{figure}

We import an image file containing the letter, map this to an $x$-$y$ coordinate system and sample in both dimensions with likelihood
\begin{align}
    \ln\L\propto\frac{-1}{xy},
\end{align}
assuming uniform priors on both variables.  Figure~\ref{fig:logo} shows the posterior distribution for the ``B'' in the \bilby{} logo.
All letters are shown in Fig~\ref{fig:logo_word}, where the axis labels have been removed.  The code for making this plot, and all other posterior distributions in the logo, are available with the {\tt git} repository~\cite{bilbygit} in {\tt \verb!sample_logo.py!}.

\section{conclusion}\label{conclusions}
Gravitational-wave astronomy is fast becoming a data-rich field.  With the significantly increased activity in the field, there is a developing need for robust, easy-to-use inference software that is also modular and adaptable.  We present \bilby: the Bayesian inference library for gravitational-wave astronomy.  \bilby{} is open-source software that can be used to perform Bayesian inference.
It is easily applied to data from LIGO/Virgo, including open data available from the Gravitational Wave Open Science Center.
We access and manipulate LIGO data using {\tt GWPy}~\cite{gwpy}.
Alternatively, \bilby{} may be used to study simulated data.
\bilby{} can also be used to perform hierarchical Bayesian inference for population studies.

We present examples highlighting \bilby's functionality and usability, including examples using open data from the first gravitational-wave detection GW150914. 
Only five lines of code are required to reconstruct the astrophysical parameters of GW150914.
One can redo the analysis using different priors, alternative waveform models, and/or a different sampling method with only modest changes.
We show how to inject binary black hole and binary neutron star signals into Monte Carlo noise.
We show how to define new gravitational-wave detectors.

We emphasise that \bilby{} is a front-end system that provides a unified interface to a variety of samplers, which are a primary workhorse of Bayesian inference.  While numerous off-the-shelf samplers are implemented (see Sec.~\ref{sec:core}), to the best of our knowledge there is no universal sampling solution to gravitational-wave parameter estimation problems.  \bilby{} is therefore only as good as the implemented samplers; initial studies show that {\tt CPNest}~\cite{veitch17}, {\tt Dynesty}~\cite{dynesty}, and {\tt emcee}~\cite{emcee,ptemcee} sample the extrinsic parameters of binary coalescences more accurately than {\tt Nestle}~\cite{nestle} and {\tt pyMultiNest}~\cite{pymultinest}.  A systematic comparison of all off-the-shelf and boutique samplers is currently underway using \bilby.

\bilby{} is designed so as to be applicable to arbitrary signal models, not just compact binary coalescences.
To this end we show two examples: one of an injected numerical relativity supernova waveform that we reconstruct using principal component analysis, and another using a proxy for a neutron star post-merger waveform.
The former example highlights how one can include their own signal models to perform both injections and signal recoveries, while the latter example demonstrates the ability to add a likelihood function that is different from the standard gravitational-wave transient likelihood.

\acknowledgments
We are grateful to John Veitch and Christopher Berry who provided valuable comments on the manuscript, and also the LIGO/Virgo Parameter Estimation group for insightful discussions.
This work is supported through Australian Research Council (ARC) Centre of Excellence CE170100004. PDL is supported through ARC Future Fellowship FT160100112 and ARC Discovery Project DP180103155.  SB is partially supported by the Australian-American Fulbright Commission.  MDP is funded by the UK Science \& Technology Facilities Council (STFC) under grant ST/N005422/1.  ET is supported through ARC Future Fellowship FT150100281 and CE170100004.
This research has made use of data, software and/or web tools obtained from the Gravitational Wave Open Science Center (https://https://www.gw-openscience.org), a service of LIGO Laboratory, the LIGO Scientific Collaboration and the Virgo Collaboration. LIGO is funded by the U.S.\ National Science Foundation. Virgo is funded by the French Centre National de Recherche Scientifique (CNRS), the Italian Istituto Nazionale della Fisica Nucleare (INFN) and the Dutch Nikhef, with contributions by Polish and Hungarian institutes.  The \bilby{} package makes use of the standard scientific \python{} stack~\citep{scipy, numpy, pandas}, {\sc matplotlib}~\cite{matplotlib}, {\sc corner}~\cite{corner}, and {\sc healpy}~\cite{healpy} for the generation of figures, {\sc deepdish}~\cite{deepdish} for {\sc hdf5}~\cite{hdf5} file operations, and {\sc astropy}~\cite{astropy1,astropy2} for common astrophysics-specific operations.

\bibliography{Bilby}

\begin{thebibliography}{103}%
\makeatletter
\providecommand \@ifxundefined [1]{%
 \@ifx{#1\undefined}
}%
\providecommand \@ifnum [1]{%
 \ifnum #1\expandafter \@firstoftwo
 \else \expandafter \@secondoftwo
 \fi
}%
\providecommand \@ifx [1]{%
 \ifx #1\expandafter \@firstoftwo
 \else \expandafter \@secondoftwo
 \fi
}%
\providecommand \natexlab [1]{#1}%
\providecommand \enquote  [1]{``#1''}%
\providecommand \bibnamefont  [1]{#1}%
\providecommand \bibfnamefont [1]{#1}%
\providecommand \citenamefont [1]{#1}%
\providecommand \href@noop [0]{\@secondoftwo}%
\providecommand \href [0]{\begingroup \@sanitize@url \@href}%
\providecommand \@href[1]{\@@startlink{#1}\@@href}%
\providecommand \@@href[1]{\endgroup#1\@@endlink}%
\providecommand \@sanitize@url [0]{\catcode `\\12\catcode `\$12\catcode
  `\&12\catcode `\#12\catcode `\^12\catcode `\_12\catcode `\%12\relax}%
\providecommand \@@startlink[1]{}%
\providecommand \@@endlink[0]{}%
\providecommand \url  [0]{\begingroup\@sanitize@url \@url }%
\providecommand \@url [1]{\endgroup\@href {#1}{\urlprefix }}%
\providecommand \urlprefix  [0]{URL }%
\providecommand \Eprint [0]{\href }%
\providecommand \doibase [0]{http://dx.doi.org/}%
\providecommand \selectlanguage [0]{\@gobble}%
\providecommand \bibinfo  [0]{\@secondoftwo}%
\providecommand \bibfield  [0]{\@secondoftwo}%
\providecommand \translation [1]{[#1]}%
\providecommand \BibitemOpen [0]{}%
\providecommand \bibitemStop [0]{}%
\providecommand \bibitemNoStop [0]{.\EOS\space}%
\providecommand \EOS [0]{\spacefactor3000\relax}%
\providecommand \BibitemShut  [1]{\csname bibitem#1\endcsname}%
\let\auto@bib@innerbib\@empty
\bibitem [{\citenamefont {{Abbott}}\ \emph
  {et~al.}(2016{\natexlab{a}})\citenamefont {{Abbott}} \emph
  {et~al.}}]{abbott16_gw150914_pe}%
  \BibitemOpen
  \bibfield  {author} {\bibinfo {author} {\bibfnamefont {B.~P.}\ \bibnamefont
  {{Abbott}}} \emph {et~al.},\ }\href {\doibase 10.1103/PhysRevLett.116.241102}
  {\bibfield  {journal} {\bibinfo  {journal} {\prl}\ }\textbf {\bibinfo
  {volume} {116}},\ \bibinfo {pages} {241102} (\bibinfo {year}
  {2016}{\natexlab{a}})}\BibitemShut {NoStop}%
\bibitem [{\citenamefont {{Abbott}}\ \emph
  {et~al.}(2016{\natexlab{b}})\citenamefont {{Abbott}} \emph
  {et~al.}}]{abbott16_gw150914_updatedpe}%
  \BibitemOpen
  \bibfield  {author} {\bibinfo {author} {\bibfnamefont {B.~P.}\ \bibnamefont
  {{Abbott}}} \emph {et~al.},\ }\href {\doibase 10.1103/PhysRevX.6.041014}
  {\bibfield  {journal} {\bibinfo  {journal} {Phys. Rev. X}\ }\textbf {\bibinfo
  {volume} {6}},\ \bibinfo {eid} {041014} (\bibinfo {year}
  {2016}{\natexlab{b}})}\BibitemShut {NoStop}%
\bibitem [{\citenamefont {{Abbott}}\ \emph
  {et~al.}(2016{\natexlab{c}})\citenamefont {{Abbott}} \emph
  {et~al.}}]{abbott16_01BBH}%
  \BibitemOpen
  \bibfield  {author} {\bibinfo {author} {\bibfnamefont {B.~P.}\ \bibnamefont
  {{Abbott}}} \emph {et~al.},\ }\href {\doibase 10.1103/PhysRevX.6.041015}
  {\bibfield  {journal} {\bibinfo  {journal} {Phys. Rev. X}\ }\textbf {\bibinfo
  {volume} {6}},\ \bibinfo {eid} {041015} (\bibinfo {year}
  {2016}{\natexlab{c}})}\BibitemShut {NoStop}%
\bibitem [{\citenamefont {{Abbott}}\ \emph
  {et~al.}(2018{\natexlab{a}})\citenamefont {{Abbott}} \emph
  {et~al.}}]{abbott18_GW170817_NS_parameters}%
  \BibitemOpen
  \bibfield  {author} {\bibinfo {author} {\bibfnamefont {B.~P.}\ \bibnamefont
  {{Abbott}}} \emph {et~al.},\ }\href@noop {} {\  (\bibinfo {year}
  {2018}{\natexlab{a}})},\ \Eprint {http://arxiv.org/abs/arXiv:1805.11581}
  {arXiv:1805.11581} \BibitemShut {NoStop}%
\bibitem [{\citenamefont {{Abbott}}\ \emph
  {et~al.}(2018{\natexlab{b}})\citenamefont {{Abbott}} \emph
  {et~al.}}]{abbott18_GW170817_properties}%
  \BibitemOpen
  \bibfield  {author} {\bibinfo {author} {\bibfnamefont {B.~P.}\ \bibnamefont
  {{Abbott}}} \emph {et~al.},\ }\href@noop {} {\  (\bibinfo {year}
  {2018}{\natexlab{b}})},\ \Eprint {http://arxiv.org/abs/arXiv:1805.11579}
  {arXiv:1805.11579} \BibitemShut {NoStop}%
\bibitem [{\citenamefont {{Abbott}}\ \emph
  {et~al.}(2017{\natexlab{a}})\citenamefont {{Abbott}} \emph
  {et~al.}}]{abbott17_gw170817_detection}%
  \BibitemOpen
  \bibfield  {author} {\bibinfo {author} {\bibfnamefont {B.~P.}\ \bibnamefont
  {{Abbott}}} \emph {et~al.},\ }\href {\doibase 10.1103/PhysRevLett.119.161101}
  {\bibfield  {journal} {\bibinfo  {journal} {\prl}\ }\textbf {\bibinfo
  {volume} {119}},\ \bibinfo {eid} {161101} (\bibinfo {year}
  {2017}{\natexlab{a}})}\BibitemShut {NoStop}%
\bibitem [{\citenamefont {{Abbott}}\ \emph
  {et~al.}(2017{\natexlab{b}})\citenamefont {{Abbott}} \emph
  {et~al.}}]{abbott17_gw170817_gwgrb}%
  \BibitemOpen
  \bibfield  {author} {\bibinfo {author} {\bibfnamefont {B.~P.}\ \bibnamefont
  {{Abbott}}} \emph {et~al.},\ }\href {\doibase 10.3847/2041-8213/aa920c}
  {\bibfield  {journal} {\bibinfo  {journal} {Astrophys. J.}\ }\textbf
  {\bibinfo {volume} {848}},\ \bibinfo {eid} {L13} (\bibinfo {year}
  {2017}{\natexlab{b}})}\BibitemShut {NoStop}%
\bibitem [{\citenamefont {{Abbott}}\ \emph
  {et~al.}(2017{\natexlab{c}})\citenamefont {{Abbott}} \emph
  {et~al.}}]{abbott17_gw170817_Hubble}%
  \BibitemOpen
  \bibfield  {author} {\bibinfo {author} {\bibfnamefont {B.~P.}\ \bibnamefont
  {{Abbott}}} \emph {et~al.},\ }\href {\doibase 10.1038/nature24471} {\bibfield
   {journal} {\bibinfo  {journal} {\nat}\ }\textbf {\bibinfo {volume} {551}},\
  \bibinfo {pages} {85} (\bibinfo {year} {2017}{\natexlab{c}})}\BibitemShut
  {NoStop}%
\bibitem [{\citenamefont {{Talbot}}\ and\ \citenamefont
  {{Thrane}}(2018)}]{talbot18}%
  \BibitemOpen
  \bibfield  {author} {\bibinfo {author} {\bibfnamefont {C.}~\bibnamefont
  {{Talbot}}}\ and\ \bibinfo {author} {\bibfnamefont {E.}~\bibnamefont
  {{Thrane}}},\ }\href {\doibase 10.3847/1538-4357/aab34c} {\bibfield
  {journal} {\bibinfo  {journal} {Astrophys. J.}\ }\textbf {\bibinfo {volume}
  {856}},\ \bibinfo {eid} {173} (\bibinfo {year} {2018})}\BibitemShut {NoStop}%
\bibitem [{\citenamefont {{Wysocki}}\ \emph {et~al.}(2018)\citenamefont
  {{Wysocki}} \emph {et~al.}}]{Wysocki2018}%
  \BibitemOpen
  \bibfield  {author} {\bibinfo {author} {\bibfnamefont {D.}~\bibnamefont
  {{Wysocki}}} \emph {et~al.},\ }\href {\doibase 10.1103/PhysRevD.97.043014}
  {\bibfield  {journal} {\bibinfo  {journal} {\prd}\ }\textbf {\bibinfo
  {volume} {97}},\ \bibinfo {eid} {043014} (\bibinfo {year}
  {2018})}\BibitemShut {NoStop}%
\bibitem [{\citenamefont {{Smith}}\ and\ \citenamefont
  {{Thrane}}(2018)}]{smith18}%
  \BibitemOpen
  \bibfield  {author} {\bibinfo {author} {\bibfnamefont {R.}~\bibnamefont
  {{Smith}}}\ and\ \bibinfo {author} {\bibfnamefont {E.}~\bibnamefont
  {{Thrane}}},\ }\href {\doibase 10.1103/PhysRevX.8.021019} {\enquote {\bibinfo
  {title} {{Optimal Search for an Astrophysical Gravitational-Wave
  Background}},}\ } (\bibinfo {year} {2018})\BibitemShut {NoStop}%
\bibitem [{\citenamefont {{Farr}}\ \emph {et~al.}(2018)\citenamefont {{Farr}},
  \citenamefont {{Holz}},\ and\ \citenamefont {{Farr}}}]{farr18}%
  \BibitemOpen
  \bibfield  {author} {\bibinfo {author} {\bibfnamefont {B.}~\bibnamefont
  {{Farr}}}, \bibinfo {author} {\bibfnamefont {D.~E.}\ \bibnamefont {{Holz}}},
  \ and\ \bibinfo {author} {\bibfnamefont {W.~M.}\ \bibnamefont {{Farr}}},\
  }\href {\doibase 10.3847/2041-8213/aaaa64} {\bibfield  {journal} {\bibinfo
  {journal} {Astrophys. J.}\ }\textbf {\bibinfo {volume} {854}},\ \bibinfo
  {eid} {L9} (\bibinfo {year} {2018})}\BibitemShut {NoStop}%
\bibitem [{\citenamefont {{Taylor}}\ and\ \citenamefont
  {{Gerosa}}(2018)}]{taylor18}%
  \BibitemOpen
  \bibfield  {author} {\bibinfo {author} {\bibfnamefont {S.~R.}\ \bibnamefont
  {{Taylor}}}\ and\ \bibinfo {author} {\bibfnamefont {D.}~\bibnamefont
  {{Gerosa}}},\ }\href@noop {} {\  (\bibinfo {year} {2018})},\ \Eprint
  {http://arxiv.org/abs/arXiv:1806.08365} {arXiv:1806.08365} \BibitemShut
  {NoStop}%
\bibitem [{\citenamefont {{Roulet}}\ and\ \citenamefont
  {{Zaldarriaga}}(2018)}]{Roulet2018}%
  \BibitemOpen
  \bibfield  {author} {\bibinfo {author} {\bibfnamefont {J.}~\bibnamefont
  {{Roulet}}}\ and\ \bibinfo {author} {\bibfnamefont {M.}~\bibnamefont
  {{Zaldarriaga}}},\ }\href@noop {} {\  (\bibinfo {year} {2018})},\ \Eprint
  {http://arxiv.org/abs/arXiv:1806.10610} {arXiv:1806.10610} \BibitemShut
  {NoStop}%
\bibitem [{\citenamefont {{Abbott}}\ \emph
  {et~al.}(2016{\natexlab{d}})\citenamefont {{Abbott}} \emph
  {et~al.}}]{abbott16_gw150914_testingGR}%
  \BibitemOpen
  \bibfield  {author} {\bibinfo {author} {\bibfnamefont {B.~P.}\ \bibnamefont
  {{Abbott}}} \emph {et~al.},\ }\href {\doibase 10.1103/PhysRevLett.116.221101}
  {\bibfield  {journal} {\bibinfo  {journal} {\prl}\ }\textbf {\bibinfo
  {volume} {116}},\ \bibinfo {eid} {221101} (\bibinfo {year}
  {2016}{\natexlab{d}})}\BibitemShut {NoStop}%
\bibitem [{\citenamefont {{Abbott}}\ \emph
  {et~al.}(2017{\natexlab{d}})\citenamefont {{Abbott}} \emph
  {et~al.}}]{abbott17_gw170814_detection}%
  \BibitemOpen
  \bibfield  {author} {\bibinfo {author} {\bibfnamefont {B.~P.}\ \bibnamefont
  {{Abbott}}} \emph {et~al.},\ }\href {\doibase 10.1103/PhysRevLett.119.141101}
  {\bibfield  {journal} {\bibinfo  {journal} {\prl}\ }\textbf {\bibinfo
  {volume} {119}},\ \bibinfo {eid} {141101} (\bibinfo {year}
  {2017}{\natexlab{d}})}\BibitemShut {NoStop}%
\bibitem [{\citenamefont {{Abbott}}\ \emph
  {et~al.}(2018{\natexlab{c}})\citenamefont {{Abbott}} \emph
  {et~al.}}]{abbott18_cw_polarisations}%
  \BibitemOpen
  \bibfield  {author} {\bibinfo {author} {\bibfnamefont {B.~P.}\ \bibnamefont
  {{Abbott}}} \emph {et~al.},\ }\href {\doibase 10.1103/PhysRevLett.120.031104}
  {\bibfield  {journal} {\bibinfo  {journal} {\prl}\ }\textbf {\bibinfo
  {volume} {120}},\ \bibinfo {eid} {031104} (\bibinfo {year}
  {2018}{\natexlab{c}})}\BibitemShut {NoStop}%
\bibitem [{\citenamefont {{Abbott}}\ \emph
  {et~al.}(2018{\natexlab{d}})\citenamefont {{Abbott}} \emph
  {et~al.}}]{abbott18_sgwb_polarisations}%
  \BibitemOpen
  \bibfield  {author} {\bibinfo {author} {\bibfnamefont {B.~P.}\ \bibnamefont
  {{Abbott}}} \emph {et~al.},\ }\href {\doibase 10.1103/PhysRevLett.120.201102}
  {\bibfield  {journal} {\bibinfo  {journal} {\prl}\ }\textbf {\bibinfo
  {volume} {120}},\ \bibinfo {eid} {201102} (\bibinfo {year}
  {2018}{\natexlab{d}})}\BibitemShut {NoStop}%
\bibitem [{\citenamefont {{Vallisneri}}\ \emph {et~al.}(2015)\citenamefont
  {{Vallisneri}}, \citenamefont {{Kanner}}, \citenamefont {{Williams}},
  \citenamefont {{Weinstein}},\ and\ \citenamefont {{Stephens}}}]{gwosc}%
  \BibitemOpen
  \bibfield  {author} {\bibinfo {author} {\bibfnamefont {M.}~\bibnamefont
  {{Vallisneri}}}, \bibinfo {author} {\bibfnamefont {J.}~\bibnamefont
  {{Kanner}}}, \bibinfo {author} {\bibfnamefont {R.}~\bibnamefont
  {{Williams}}}, \bibinfo {author} {\bibfnamefont {A.}~\bibnamefont
  {{Weinstein}}}, \ and\ \bibinfo {author} {\bibfnamefont {B.}~\bibnamefont
  {{Stephens}}},\ }in\ \href {\doibase 10.1088/1742-6596/610/1/012021} {\emph
  {\bibinfo {booktitle} {Journal of Physics Conference Series}}},\ \bibinfo
  {series} {Journal of Physics Conference Series}, Vol.\ \bibinfo {volume}
  {610}\ (\bibinfo {year} {2015})\ p.\ \bibinfo {pages} {012021}\BibitemShut
  {NoStop}%
\bibitem [{\citenamefont {{Veitch}}\ \emph {et~al.}(2015)\citenamefont
  {{Veitch}} \emph {et~al.}}]{veitch15}%
  \BibitemOpen
  \bibfield  {author} {\bibinfo {author} {\bibfnamefont {J.}~\bibnamefont
  {{Veitch}}} \emph {et~al.},\ }\href {\doibase 10.1103/PhysRevD.91.042003}
  {\bibfield  {journal} {\bibinfo  {journal} {\prd}\ }\textbf {\bibinfo
  {volume} {91}},\ \bibinfo {eid} {042003} (\bibinfo {year}
  {2015})}\BibitemShut {NoStop}%
\bibitem [{\citenamefont {{Biwer}}\ \emph {et~al.}(2018)\citenamefont
  {{Biwer}}, \citenamefont {{Capano}}, \citenamefont {{De}}, \citenamefont
  {{Cabero}}, \citenamefont {{Brown}}, \citenamefont {{Nitz}},\ and\
  \citenamefont {{Raymond}}}]{biwer18}%
  \BibitemOpen
  \bibfield  {author} {\bibinfo {author} {\bibfnamefont {C.~M.}\ \bibnamefont
  {{Biwer}}}, \bibinfo {author} {\bibfnamefont {C.~D.}\ \bibnamefont
  {{Capano}}}, \bibinfo {author} {\bibfnamefont {S.}~\bibnamefont {{De}}},
  \bibinfo {author} {\bibfnamefont {M.}~\bibnamefont {{Cabero}}}, \bibinfo
  {author} {\bibfnamefont {D.~A.}\ \bibnamefont {{Brown}}}, \bibinfo {author}
  {\bibfnamefont {A.~H.}\ \bibnamefont {{Nitz}}}, \ and\ \bibinfo {author}
  {\bibfnamefont {V.}~\bibnamefont {{Raymond}}},\ }\href@noop {} {\  (\bibinfo
  {year} {2018})},\ \Eprint {http://arxiv.org/abs/arXiv:1807.10312}
  {arXiv:1807.10312} \BibitemShut {NoStop}%
\bibitem [{\citenamefont {Nitz}\ \emph {et~al.}(2018)\citenamefont {Nitz},
  \citenamefont {Harry}, \citenamefont {Brown}, \citenamefont {Biwer} \emph
  {et~al.}}]{nitz18}%
  \BibitemOpen
  \bibfield  {author} {\bibinfo {author} {\bibfnamefont {A.}~\bibnamefont
  {Nitz}}, \bibinfo {author} {\bibfnamefont {I.}~\bibnamefont {Harry}},
  \bibinfo {author} {\bibfnamefont {D.}~\bibnamefont {Brown}}, \bibinfo
  {author} {\bibfnamefont {C.~M.}\ \bibnamefont {Biwer}},  \emph {et~al.},\
  }\href {\doibase 10.5281/zenodo.1410598} {\enquote {\bibinfo {title}
  {gwastro/pycbc: 1.12.3 release},}\ } (\bibinfo {year} {2018})\BibitemShut
  {NoStop}%
\bibitem [{\citenamefont {{Abbott}}\ \emph
  {et~al.}(2016{\natexlab{e}})\citenamefont {{Abbott}} \emph
  {et~al.}}]{abbott16_gw150914_detection}%
  \BibitemOpen
  \bibfield  {author} {\bibinfo {author} {\bibfnamefont {B.~P.}\ \bibnamefont
  {{Abbott}}} \emph {et~al.},\ }\href {\doibase 10.1103/PhysRevLett.116.061102}
  {\bibfield  {journal} {\bibinfo  {journal} {\prl}\ }\textbf {\bibinfo
  {volume} {116}},\ \bibinfo {pages} {061102} (\bibinfo {year}
  {2016}{\natexlab{e}})}\BibitemShut {NoStop}%
\bibitem [{\citenamefont {{Abbott}}\ \emph
  {et~al.}(2016{\natexlab{f}})\citenamefont {{Abbott}} \emph
  {et~al.}}]{abbott16_gw151226_detection}%
  \BibitemOpen
  \bibfield  {author} {\bibinfo {author} {\bibfnamefont {B.~P.}\ \bibnamefont
  {{Abbott}}} \emph {et~al.},\ }\href {\doibase 10.1103/PhysRevLett.116.241103}
  {\bibfield  {journal} {\bibinfo  {journal} {\prl}\ }\textbf {\bibinfo
  {volume} {116}},\ \bibinfo {pages} {241103} (\bibinfo {year}
  {2016}{\natexlab{f}})}\BibitemShut {NoStop}%
\bibitem [{\citenamefont {{Abbott}}\ \emph
  {et~al.}(2017{\natexlab{e}})\citenamefont {{Abbott}} \emph
  {et~al.}}]{abbott17_gw170104_detection}%
  \BibitemOpen
  \bibfield  {author} {\bibinfo {author} {\bibfnamefont {B.~P.}\ \bibnamefont
  {{Abbott}}} \emph {et~al.},\ }\href {\doibase 10.1103/PhysRevLett.118.221101}
  {\bibfield  {journal} {\bibinfo  {journal} {\prl}\ }\textbf {\bibinfo
  {volume} {118}},\ \bibinfo {eid} {221101} (\bibinfo {year}
  {2017}{\natexlab{e}})}\BibitemShut {NoStop}%
\bibitem [{\citenamefont {{Abbott}}\ \emph
  {et~al.}(2017{\natexlab{f}})\citenamefont {{Abbott}} \emph
  {et~al.}}]{abbott17_gw170608_detection}%
  \BibitemOpen
  \bibfield  {author} {\bibinfo {author} {\bibfnamefont {B.~P.}\ \bibnamefont
  {{Abbott}}} \emph {et~al.},\ }\href {\doibase 10.3847/2041-8213/aa9f0c}
  {\bibfield  {journal} {\bibinfo  {journal} {Astrophys. J.}\ }\textbf
  {\bibinfo {volume} {851}},\ \bibinfo {eid} {L35} (\bibinfo {year}
  {2017}{\natexlab{f}})}\BibitemShut {NoStop}%
\bibitem [{\citenamefont {Skilling}(2004)}]{Skilling04}%
  \BibitemOpen
  \bibfield  {author} {\bibinfo {author} {\bibfnamefont {J.}~\bibnamefont
  {Skilling}},\ }\href {\doibase 10.1063/1.1835238} {\bibfield  {journal}
  {\bibinfo  {journal} {AIP Conf. Proc.}\ }\textbf {\bibinfo {volume} {735}},\
  \bibinfo {pages} {395} (\bibinfo {year} {2004})}\BibitemShut {NoStop}%
\bibitem [{\citenamefont {{Thrane}}\ and\ \citenamefont
  {{Talbot}}(2018)}]{thrane18}%
  \BibitemOpen
  \bibfield  {author} {\bibinfo {author} {\bibfnamefont {E.}~\bibnamefont
  {{Thrane}}}\ and\ \bibinfo {author} {\bibfnamefont {C.}~\bibnamefont
  {{Talbot}}},\ }\href@noop {} {\enquote {\bibinfo {title} {{An introduction to
  Bayesian inference in gravitational-wave astronomy: parameter estimation,
  model selection, and hierarchical models}},}\ } (\bibinfo {year} {2018}),\
  \Eprint {http://arxiv.org/abs/arXiv:1809.02293} {arXiv:1809.02293}
  \BibitemShut {NoStop}%
\bibitem [{\citenamefont {Pierce}(2002)}]{pierce02}%
  \BibitemOpen
  \bibfield  {author} {\bibinfo {author} {\bibfnamefont {B.}~\bibnamefont
  {Pierce}},\ }\href@noop {} {\emph {\bibinfo {title} {Types and Programming
  Languages}}}\ (\bibinfo  {publisher} {The MIT Press},\ \bibinfo {year}
  {2002})\BibitemShut {NoStop}%
\bibitem [{\citenamefont {{Buchner}}\ \emph {et~al.}(2014)\citenamefont
  {{Buchner}} \emph {et~al.}}]{pymultinest}%
  \BibitemOpen
  \bibfield  {author} {\bibinfo {author} {\bibfnamefont {J.}~\bibnamefont
  {{Buchner}}} \emph {et~al.},\ }\href {\doibase 10.1051/0004-6361/201322971}
  {\bibfield  {journal} {\bibinfo  {journal} {Astron. Astrophys.}\ }\textbf
  {\bibinfo {volume} {564}},\ \bibinfo {eid} {A125} (\bibinfo {year}
  {2014})}\BibitemShut {NoStop}%
\bibitem [{\citenamefont {{Foreman-Mackey}}\ \emph {et~al.}(2013)\citenamefont
  {{Foreman-Mackey}}, \citenamefont {{Hogg}}, \citenamefont {{Lang}},\ and\
  \citenamefont {{Goodman}}}]{emcee}%
  \BibitemOpen
  \bibfield  {author} {\bibinfo {author} {\bibfnamefont {D.}~\bibnamefont
  {{Foreman-Mackey}}}, \bibinfo {author} {\bibfnamefont {D.~W.}\ \bibnamefont
  {{Hogg}}}, \bibinfo {author} {\bibfnamefont {D.}~\bibnamefont {{Lang}}}, \
  and\ \bibinfo {author} {\bibfnamefont {J.}~\bibnamefont {{Goodman}}},\ }\href
  {\doibase 10.1086/670067} {\bibfield  {journal} {\bibinfo  {journal} {Publ.
  Astron. Soc. Pac.}\ }\textbf {\bibinfo {volume} {125}},\ \bibinfo {pages}
  {306} (\bibinfo {year} {2013})}\BibitemShut {NoStop}%
\bibitem [{Note1()}]{Note1}%
  \BibitemOpen
  \bibinfo {note} {\protect \url
  {https://www.python.org/dev/peps/pep-0008/}}\BibitemShut {NoStop}%
\bibitem [{Note2()}]{Note2}%
  \BibitemOpen
  \bibinfo {note} {\protect \url
  {https://www.python.org/dev/peps/pep-0020/}}\BibitemShut {NoStop}%
\bibitem [{Note3()}]{Note3}%
  \BibitemOpen
  \bibinfo {note} {\protect \url {https://pypi.org/project/BILBY/}}\BibitemShut
  {NoStop}%
\bibitem [{bil({\natexlab{a}})}]{bilbygit}%
  \BibitemOpen
  \href@noop {} {}\bibinfo {howpublished}
  {\url{https://git.ligo.org/lscsoft/bilby/}} ({\natexlab{a}})\BibitemShut
  {NoStop}%
\bibitem [{bil({\natexlab{b}})}]{bilbydoc}%
  \BibitemOpen
  \href@noop {} {}\bibinfo {howpublished}
  {\url{https://lscsoft.docs.ligo.org/bilby/}} ({\natexlab{b}})\BibitemShut
  {NoStop}%
\bibitem [{\citenamefont {{Vousden}}\ \emph {et~al.}(2016)\citenamefont
  {{Vousden}}, \citenamefont {{Farr}},\ and\ \citenamefont
  {{Mandel}}}]{ptemcee}%
  \BibitemOpen
  \bibfield  {author} {\bibinfo {author} {\bibfnamefont {W.~D.}\ \bibnamefont
  {{Vousden}}}, \bibinfo {author} {\bibfnamefont {W.~M.}\ \bibnamefont
  {{Farr}}}, \ and\ \bibinfo {author} {\bibfnamefont {I.}~\bibnamefont
  {{Mandel}}},\ }\href {\doibase 10.1093/mnras/stv2422} {\bibfield  {journal}
  {\bibinfo  {journal} {Mon. Not. R. Astron. Soc.}\ }\textbf {\bibinfo {volume}
  {455}},\ \bibinfo {pages} {1919} (\bibinfo {year} {2016})}\BibitemShut
  {NoStop}%
\bibitem [{\citenamefont {Salvatier}\ \emph {et~al.}(2016)\citenamefont
  {Salvatier}, \citenamefont {Wiecki},\ and\ \citenamefont
  {Fonnesbeck}}]{salvatier16}%
  \BibitemOpen
  \bibfield  {author} {\bibinfo {author} {\bibfnamefont {J.}~\bibnamefont
  {Salvatier}}, \bibinfo {author} {\bibfnamefont {T.}~\bibnamefont {Wiecki}}, \
  and\ \bibinfo {author} {\bibfnamefont {C.}~\bibnamefont {Fonnesbeck}},\
  }\href {\doibase 10.7717/peerj-cs.55} {\bibfield  {journal} {\bibinfo
  {journal} {PeerJ Computer Science}\ }\textbf {\bibinfo {volume} {2}},\
  \bibinfo {pages} {55} (\bibinfo {year} {2016})}\BibitemShut {NoStop}%
\bibitem [{\citenamefont {Skilling}(2006)}]{Skilling06}%
  \BibitemOpen
  \bibfield  {author} {\bibinfo {author} {\bibfnamefont {J.}~\bibnamefont
  {Skilling}},\ }\href {\doibase 10.1214/06-BA127} {\bibfield  {journal}
  {\bibinfo  {journal} {Bayesian Anal.}\ }\textbf {\bibinfo {volume} {1}},\
  \bibinfo {pages} {833} (\bibinfo {year} {2006})}\BibitemShut {NoStop}%
\bibitem [{\citenamefont {{Feroz}}\ and\ \citenamefont
  {{Hobson}}(2008)}]{multinest1}%
  \BibitemOpen
  \bibfield  {author} {\bibinfo {author} {\bibfnamefont {F.}~\bibnamefont
  {{Feroz}}}\ and\ \bibinfo {author} {\bibfnamefont {M.~P.}\ \bibnamefont
  {{Hobson}}},\ }\href {\doibase 10.1111/j.1365-2966.2007.12353.x} {\bibfield
  {journal} {\bibinfo  {journal} {Mon. Not. R. Astron. Soc.}\ }\textbf
  {\bibinfo {volume} {384}},\ \bibinfo {pages} {449} (\bibinfo {year}
  {2008})}\BibitemShut {NoStop}%
\bibitem [{\citenamefont {{Feroz}}\ \emph {et~al.}(2009)\citenamefont
  {{Feroz}}, \citenamefont {{Hobson}},\ and\ \citenamefont
  {{Bridges}}}]{multinest2}%
  \BibitemOpen
  \bibfield  {author} {\bibinfo {author} {\bibfnamefont {F.}~\bibnamefont
  {{Feroz}}}, \bibinfo {author} {\bibfnamefont {M.~P.}\ \bibnamefont
  {{Hobson}}}, \ and\ \bibinfo {author} {\bibfnamefont {M.}~\bibnamefont
  {{Bridges}}},\ }\href {\doibase 10.1111/j.1365-2966.2009.14548.x} {\bibfield
  {journal} {\bibinfo  {journal} {Mon. Not. R. Astron. Soc.}\ }\textbf
  {\bibinfo {volume} {398}},\ \bibinfo {pages} {1601} (\bibinfo {year}
  {2009})}\BibitemShut {NoStop}%
\bibitem [{\citenamefont {{Feroz}}\ \emph {et~al.}(2013)\citenamefont
  {{Feroz}}, \citenamefont {{Hobson}}, \citenamefont {{Cameron}},\ and\
  \citenamefont {{Pettitt}}}]{multinest3}%
  \BibitemOpen
  \bibfield  {author} {\bibinfo {author} {\bibfnamefont {F.}~\bibnamefont
  {{Feroz}}}, \bibinfo {author} {\bibfnamefont {M.~P.}\ \bibnamefont
  {{Hobson}}}, \bibinfo {author} {\bibfnamefont {E.}~\bibnamefont {{Cameron}}},
  \ and\ \bibinfo {author} {\bibfnamefont {A.~N.}\ \bibnamefont {{Pettitt}}},\
  }\href@noop {} {\  (\bibinfo {year} {2013})},\ \Eprint
  {http://arxiv.org/abs/arXiv:1306.2144} {arXiv:1306.2144} \BibitemShut
  {NoStop}%
\bibitem [{nes()}]{nestle}%
  \BibitemOpen
  \href@noop {} {}\bibinfo {howpublished}
  {\url{http://kylebarbary.com/nestle/}}\BibitemShut {NoStop}%
\bibitem [{dyn()}]{dynesty}%
  \BibitemOpen
  \href@noop {} {}\bibinfo {howpublished}
  {\url{https://github.com/joshspeagle/dynesty}}\BibitemShut {NoStop}%
\bibitem [{\citenamefont {Veitch}\ \emph {et~al.}(2017)\citenamefont {Veitch}
  \emph {et~al.}}]{veitch17}%
  \BibitemOpen
  \bibfield  {author} {\bibinfo {author} {\bibfnamefont {J.}~\bibnamefont
  {Veitch}} \emph {et~al.},\ }\href {\doibase 10.5281/zenodo.835874} {}\bibinfo
  {howpublished} {\url{https://github.com/johnveitch/cpnest}} (\bibinfo {year}
  {2017})\BibitemShut {NoStop}%
\bibitem [{\citenamefont {{The HDF Group}}(NNNN)}]{hdf5}%
  \BibitemOpen
  \bibfield  {author} {\bibinfo {author} {\bibnamefont {{The HDF Group}}},\
  }\href@noop {} {\enquote {\bibinfo {title} {{Hierarchical Data Format,
  version 5}},}\ } (\bibinfo {year} {1997-NNNN}),\ \bibinfo {note}
  {http://www.hdfgroup.org/HDF5/}\BibitemShut {NoStop}%
\bibitem [{\citenamefont {LSC}()}]{LALSuite}%
  \BibitemOpen
  \bibfield  {author} {\bibinfo {author} {\bibnamefont {LSC}},\ }\href@noop {}
  {\enquote {\bibinfo {title} {{LIGO Scientific Collaboration Algorithm Library
  software packages: LALInference, LALSimulation, and LALInspiral}},}\
  }\bibinfo {howpublished}
  {\url{https://wiki.ligo.org/DASWG/LALSuite}}\BibitemShut {NoStop}%
\bibitem [{\citenamefont {Macleod}\ \emph {et~al.}(2018)\citenamefont
  {Macleod}, \citenamefont {Coughlin}, \citenamefont {Urban}, \citenamefont
  {Massinger} \emph {et~al.}}]{gwpy}%
  \BibitemOpen
  \bibfield  {author} {\bibinfo {author} {\bibfnamefont {D.}~\bibnamefont
  {Macleod}}, \bibinfo {author} {\bibfnamefont {S.}~\bibnamefont {Coughlin}},
  \bibinfo {author} {\bibfnamefont {A.~L.}\ \bibnamefont {Urban}}, \bibinfo
  {author} {\bibfnamefont {T.}~\bibnamefont {Massinger}},  \emph {et~al.},\
  }\href {\doibase 10.5281/zenodo.1346349} {\  (\bibinfo {year} {2018}),\
  10.5281/zenodo.1346349}\BibitemShut {NoStop}%
\bibitem [{\citenamefont {{van der Sluys}}\ \emph
  {et~al.}(2008{\natexlab{a}})\citenamefont {{van der Sluys}}, \citenamefont
  {{Raymond}}, \citenamefont {{Mandel}}, \citenamefont {{R{\"o}ver}},
  \citenamefont {{Christensen}}, \citenamefont {{Kalogera}}, \citenamefont
  {{Meyer}},\ and\ \citenamefont {{Vecchio}}}]{vandersluys08a}%
  \BibitemOpen
  \bibfield  {author} {\bibinfo {author} {\bibfnamefont {M.}~\bibnamefont {{van
  der Sluys}}}, \bibinfo {author} {\bibfnamefont {V.}~\bibnamefont
  {{Raymond}}}, \bibinfo {author} {\bibfnamefont {I.}~\bibnamefont {{Mandel}}},
  \bibinfo {author} {\bibfnamefont {C.}~\bibnamefont {{R{\"o}ver}}}, \bibinfo
  {author} {\bibfnamefont {N.}~\bibnamefont {{Christensen}}}, \bibinfo {author}
  {\bibfnamefont {V.}~\bibnamefont {{Kalogera}}}, \bibinfo {author}
  {\bibfnamefont {R.}~\bibnamefont {{Meyer}}}, \ and\ \bibinfo {author}
  {\bibfnamefont {A.}~\bibnamefont {{Vecchio}}},\ }\href {\doibase
  10.1088/0264-9381/25/18/184011} {\bibfield  {journal} {\bibinfo  {journal}
  {Classical and Quantum Gravity}\ }\textbf {\bibinfo {volume} {25}},\ \bibinfo
  {eid} {184011} (\bibinfo {year} {2008}{\natexlab{a}})}\BibitemShut {NoStop}%
\bibitem [{\citenamefont {{van der Sluys}}\ \emph
  {et~al.}(2008{\natexlab{b}})\citenamefont {{van der Sluys}}, \citenamefont
  {{R{\"o}ver}}, \citenamefont {{Stroeer}}, \citenamefont {{Raymond}},
  \citenamefont {{Mandel}}, \citenamefont {{Christensen}}, \citenamefont
  {{Kalogera}}, \citenamefont {{Meyer}},\ and\ \citenamefont
  {{Vecchio}}}]{vandersluys08b}%
  \BibitemOpen
  \bibfield  {author} {\bibinfo {author} {\bibfnamefont {M.~V.}\ \bibnamefont
  {{van der Sluys}}}, \bibinfo {author} {\bibfnamefont {C.}~\bibnamefont
  {{R{\"o}ver}}}, \bibinfo {author} {\bibfnamefont {A.}~\bibnamefont
  {{Stroeer}}}, \bibinfo {author} {\bibfnamefont {V.}~\bibnamefont
  {{Raymond}}}, \bibinfo {author} {\bibfnamefont {I.}~\bibnamefont {{Mandel}}},
  \bibinfo {author} {\bibfnamefont {N.}~\bibnamefont {{Christensen}}}, \bibinfo
  {author} {\bibfnamefont {V.}~\bibnamefont {{Kalogera}}}, \bibinfo {author}
  {\bibfnamefont {R.}~\bibnamefont {{Meyer}}}, \ and\ \bibinfo {author}
  {\bibfnamefont {A.}~\bibnamefont {{Vecchio}}},\ }\href {\doibase
  10.1086/595279} {\bibfield  {journal} {\bibinfo  {journal} {\apjl}\ }\textbf
  {\bibinfo {volume} {688}},\ \bibinfo {pages} {L61} (\bibinfo {year}
  {2008}{\natexlab{b}})},\ \Eprint {http://arxiv.org/abs/0710.1897}
  {arXiv:0710.1897} \BibitemShut {NoStop}%
\bibitem [{\citenamefont {{Veitch}}\ and\ \citenamefont
  {{Vecchio}}(2008)}]{veitch08}%
  \BibitemOpen
  \bibfield  {author} {\bibinfo {author} {\bibfnamefont {J.}~\bibnamefont
  {{Veitch}}}\ and\ \bibinfo {author} {\bibfnamefont {A.}~\bibnamefont
  {{Vecchio}}},\ }\href {\doibase 10.1103/PhysRevD.78.022001} {\bibfield
  {journal} {\bibinfo  {journal} {\prd}\ }\textbf {\bibinfo {volume} {78}},\
  \bibinfo {eid} {022001} (\bibinfo {year} {2008})}\BibitemShut {NoStop}%
\bibitem [{\citenamefont {{Schmidt}}\ \emph {et~al.}(2012)\citenamefont
  {{Schmidt}}, \citenamefont {{Hannam}},\ and\ \citenamefont
  {{Husa}}}]{IMRPhenomP}%
  \BibitemOpen
  \bibfield  {author} {\bibinfo {author} {\bibfnamefont {P.}~\bibnamefont
  {{Schmidt}}}, \bibinfo {author} {\bibfnamefont {M.}~\bibnamefont {{Hannam}}},
  \ and\ \bibinfo {author} {\bibfnamefont {S.}~\bibnamefont {{Husa}}},\
  }\href@noop {} {\bibfield  {journal} {\bibinfo  {journal} {\prd}\ }\textbf
  {\bibinfo {volume} {86}},\ \bibinfo {eid} {104063} (\bibinfo {year}
  {2012})}\BibitemShut {NoStop}%
\bibitem [{\citenamefont {{Aasi}}\ \emph {et~al.}(2015)\citenamefont {{Aasi}}
  \emph {et~al.}}]{LIGO}%
  \BibitemOpen
  \bibfield  {author} {\bibinfo {author} {\bibfnamefont {J.}~\bibnamefont
  {{Aasi}}} \emph {et~al.},\ }\href {\doibase 10.1088/0264-9381/32/7/074001}
  {\bibfield  {journal} {\bibinfo  {journal} {Classical Quantum Gravity}\
  }\textbf {\bibinfo {volume} {32}},\ \bibinfo {eid} {074001} (\bibinfo {year}
  {2015})}\BibitemShut {NoStop}%
\bibitem [{\citenamefont {{Raymond}}\ and\ \citenamefont
  {{Farr}}(2014)}]{raymond14}%
  \BibitemOpen
  \bibfield  {author} {\bibinfo {author} {\bibfnamefont {V.}~\bibnamefont
  {{Raymond}}}\ and\ \bibinfo {author} {\bibfnamefont {W.~M.}\ \bibnamefont
  {{Farr}}},\ }\href@noop {} {\  (\bibinfo {year} {2014})},\ \Eprint
  {http://arxiv.org/abs/arXiv:1402.0053} {arXiv:1402.0053} \BibitemShut
  {NoStop}%
\bibitem [{\citenamefont {{Farr}}\ \emph
  {et~al.}(2014{\natexlab{a}})\citenamefont {{Farr}}, \citenamefont
  {{Ochsner}}, \citenamefont {{Farr}},\ and\ \citenamefont
  {{O'Shaughnessy}}}]{farr14c}%
  \BibitemOpen
  \bibfield  {author} {\bibinfo {author} {\bibfnamefont {B.}~\bibnamefont
  {{Farr}}}, \bibinfo {author} {\bibfnamefont {E.}~\bibnamefont {{Ochsner}}},
  \bibinfo {author} {\bibfnamefont {W.~M.}\ \bibnamefont {{Farr}}}, \ and\
  \bibinfo {author} {\bibfnamefont {R.}~\bibnamefont {{O'Shaughnessy}}},\
  }\href {\doibase 10.1103/PhysRevD.90.024018} {\bibfield  {journal} {\bibinfo
  {journal} {\prd}\ }\textbf {\bibinfo {volume} {90}},\ \bibinfo {eid} {024018}
  (\bibinfo {year} {2014}{\natexlab{a}})}\BibitemShut {NoStop}%
\bibitem [{\citenamefont {{Farr}}(2014)}]{farr14}%
  \BibitemOpen
  \bibfield  {author} {\bibinfo {author} {\bibfnamefont {W.~M.}\ \bibnamefont
  {{Farr}}},\ }\href@noop {} {\enquote {\bibinfo {title} {Marginalisation of
  the time and phase parameters in cbc parameter estimation},}\ } (\bibinfo
  {year} {2014}),\ \bibinfo {note}
  {\url{https://dcc.ligo.org/T1400460-v2/public}}\BibitemShut {NoStop}%
\bibitem [{\citenamefont {{Singer}}\ and\ \citenamefont
  {{Price}}(2016)}]{singer16a}%
  \BibitemOpen
  \bibfield  {author} {\bibinfo {author} {\bibfnamefont {L.~P.}\ \bibnamefont
  {{Singer}}}\ and\ \bibinfo {author} {\bibfnamefont {L.~R.}\ \bibnamefont
  {{Price}}},\ }\href {\doibase 10.1103/PhysRevD.93.024013} {\bibfield
  {journal} {\bibinfo  {journal} {\prd}\ }\textbf {\bibinfo {volume} {93}},\
  \bibinfo {eid} {024013} (\bibinfo {year} {2016})}\BibitemShut {NoStop}%
\bibitem [{\citenamefont {{Singer}}\ \emph {et~al.}(2016)\citenamefont
  {{Singer}} \emph {et~al.}}]{singer16b}%
  \BibitemOpen
  \bibfield  {author} {\bibinfo {author} {\bibfnamefont {L.~P.}\ \bibnamefont
  {{Singer}}} \emph {et~al.},\ }\href {\doibase 10.3847/2041-8205/829/1/L15}
  {\bibfield  {journal} {\bibinfo  {journal} {Astrophys. J.}\ }\textbf
  {\bibinfo {volume} {829}},\ \bibinfo {eid} {L15} (\bibinfo {year}
  {2016})}\BibitemShut {NoStop}%
\bibitem [{\citenamefont {{Farr}}\ \emph
  {et~al.}(2014{\natexlab{b}})\citenamefont {{Farr}}, \citenamefont {Farr},\
  and\ \citenamefont {Littenberg}}]{farr14b}%
  \BibitemOpen
  \bibfield  {author} {\bibinfo {author} {\bibfnamefont {W.~M.}\ \bibnamefont
  {{Farr}}}, \bibinfo {author} {\bibfnamefont {B.}~\bibnamefont {Farr}}, \ and\
  \bibinfo {author} {\bibfnamefont {T.}~\bibnamefont {Littenberg}},\
  }\href@noop {} {\enquote {\bibinfo {title} {Modelling calibration errors in
  cbc waveforms},}\ } (\bibinfo {year} {2014}{\natexlab{b}}),\ \bibinfo {note}
  {\url{https://dcc.ligo.org/LIGO-T1400682/public}}\BibitemShut {NoStop}%
\bibitem [{Note4()}]{Note4}%
  \BibitemOpen
  \bibinfo {note} {This example is found in the {\protect \sc Bilby}\protect
  \xspace {} {\protect \tt git} repository at \protect \url
  {https://git.ligo.org/lscsoft/bilby/blob/master/examples/injection_examples/basic_tutorial.py}.}\BibitemShut
  {Stop}%
\bibitem [{\citenamefont {{Acernese}}\ \emph {et~al.}(2015)\citenamefont
  {{Acernese}} \emph {et~al.}}]{virgo}%
  \BibitemOpen
  \bibfield  {author} {\bibinfo {author} {\bibfnamefont {F.}~\bibnamefont
  {{Acernese}}} \emph {et~al.},\ }\href {\doibase
  10.1088/0264-9381/32/2/024001} {\bibfield  {journal} {\bibinfo  {journal}
  {Classical Quantum Gravity}\ }\textbf {\bibinfo {volume} {32}},\ \bibinfo
  {eid} {024001} (\bibinfo {year} {2015})}\BibitemShut {NoStop}%
\bibitem [{\citenamefont {{Abbott}}\ \emph
  {et~al.}(2017{\natexlab{g}})\citenamefont {{Abbott}} \emph
  {et~al.}}]{abbott17_gw170817_multimessenger}%
  \BibitemOpen
  \bibfield  {author} {\bibinfo {author} {\bibfnamefont {B.~P.}\ \bibnamefont
  {{Abbott}}} \emph {et~al.},\ }\href {\doibase 10.3847/2041-8213/aa91c9}
  {\bibfield  {journal} {\bibinfo  {journal} {Astrophys. J.}\ }\textbf
  {\bibinfo {volume} {848}},\ \bibinfo {eid} {L12} (\bibinfo {year}
  {2017}{\natexlab{g}})}\BibitemShut {NoStop}%
\bibitem [{\citenamefont {{De}}\ \emph {et~al.}(2018)\citenamefont {{De}},
  \citenamefont {{Finstad}}, \citenamefont {{Lattimer}}, \citenamefont
  {{Brown}}, \citenamefont {{Berger}},\ and\ \citenamefont {{Biwer}}}]{de18}%
  \BibitemOpen
  \bibfield  {author} {\bibinfo {author} {\bibfnamefont {S.}~\bibnamefont
  {{De}}}, \bibinfo {author} {\bibfnamefont {D.}~\bibnamefont {{Finstad}}},
  \bibinfo {author} {\bibfnamefont {J.~M.}\ \bibnamefont {{Lattimer}}},
  \bibinfo {author} {\bibfnamefont {D.~A.}\ \bibnamefont {{Brown}}}, \bibinfo
  {author} {\bibfnamefont {E.}~\bibnamefont {{Berger}}}, \ and\ \bibinfo
  {author} {\bibfnamefont {C.~M.}\ \bibnamefont {{Biwer}}},\ }\href@noop {} {\
  (\bibinfo {year} {2018})},\ \Eprint {http://arxiv.org/abs/arXiv:1804.08583}
  {arXiv:1804.08583} \BibitemShut {NoStop}%
\bibitem [{Note5()}]{Note5}%
  \BibitemOpen
  \bibinfo {note} {This example is found in the {\protect \sc Bilby}\protect
  \xspace {} {\protect \tt git} repository at \protect \url
  {https://git.ligo.org/lscsoft/bilby/blob/master/examples/injection_examples/binary_neutron_star_example.py}.}\BibitemShut
  {Stop}%
\bibitem [{\citenamefont {{Flanagan}}\ and\ \citenamefont
  {{Hinderer}}(2008)}]{flanagan08}%
  \BibitemOpen
  \bibfield  {author} {\bibinfo {author} {\bibfnamefont {{\'E}.~{\'E}.}\
  \bibnamefont {{Flanagan}}}\ and\ \bibinfo {author} {\bibfnamefont
  {T.}~\bibnamefont {{Hinderer}}},\ }\href {\doibase
  10.1103/PhysRevD.77.021502} {\bibfield  {journal} {\bibinfo  {journal}
  {\prd}\ }\textbf {\bibinfo {volume} {77}},\ \bibinfo {eid} {021502} (\bibinfo
  {year} {2008})}\BibitemShut {NoStop}%
\bibitem [{\citenamefont {{Lackey}}\ and\ \citenamefont
  {{Wade}}(2015)}]{lackey15}%
  \BibitemOpen
  \bibfield  {author} {\bibinfo {author} {\bibfnamefont {B.~D.}\ \bibnamefont
  {{Lackey}}}\ and\ \bibinfo {author} {\bibfnamefont {L.}~\bibnamefont
  {{Wade}}},\ }\href@noop {} {\bibfield  {journal} {\bibinfo  {journal} {\prd}\
  }\textbf {\bibinfo {volume} {91}},\ \bibinfo {eid} {043002} (\bibinfo {year}
  {2015})}\BibitemShut {NoStop}%
\bibitem [{\citenamefont {{Lower}}\ \emph {et~al.}(2018)\citenamefont
  {{Lower}}, \citenamefont {{Thrane}}, \citenamefont {{Lasky}},\ and\
  \citenamefont {{Smith}}}]{lower18}%
  \BibitemOpen
  \bibfield  {author} {\bibinfo {author} {\bibfnamefont {M.~E.}\ \bibnamefont
  {{Lower}}}, \bibinfo {author} {\bibfnamefont {E.}~\bibnamefont {{Thrane}}},
  \bibinfo {author} {\bibfnamefont {P.~D.}\ \bibnamefont {{Lasky}}}, \ and\
  \bibinfo {author} {\bibfnamefont {R.}~\bibnamefont {{Smith}}},\ }\href
  {\doibase 10.1103/PhysRevD.98.083028} {\bibfield  {journal} {\bibinfo
  {journal} {Phys. Rev. D}\ }\textbf {\bibinfo {volume} {98}},\ \bibinfo
  {pages} {083028} (\bibinfo {year} {2018})}\BibitemShut {NoStop}%
\bibitem [{\citenamefont {Iyer}\ \emph {et~al.}(2011)\citenamefont {Iyer},
  \citenamefont {Souradeep}, \citenamefont {Unnikrishnan}, \citenamefont
  {Dhurandhar}, \citenamefont {Raja}, \citenamefont {Sengupta},\ and\
  \citenamefont {Consortium)}}]{LIGOIndia}%
  \BibitemOpen
  \bibfield  {author} {\bibinfo {author} {\bibfnamefont {B.~R.}\ \bibnamefont
  {Iyer}}, \bibinfo {author} {\bibfnamefont {T.}~\bibnamefont {Souradeep}},
  \bibinfo {author} {\bibfnamefont {C.~S.}\ \bibnamefont {Unnikrishnan}},
  \bibinfo {author} {\bibfnamefont {S.}~\bibnamefont {Dhurandhar}}, \bibinfo
  {author} {\bibfnamefont {S.}~\bibnamefont {Raja}}, \bibinfo {author}
  {\bibfnamefont {A.}~\bibnamefont {Sengupta}}, \ and\ \bibinfo {author}
  {\bibfnamefont {I.}~\bibnamefont {Consortium)}},\ }\href@noop {} {\enquote
  {\bibinfo {title} {Ligo-india, proposal of the consortium for indian
  initiative in gravitational-wave observations (indigo)},}\ } (\bibinfo {year}
  {2011}),\ \bibinfo {note}
  {\url{https://dcc.ligo.org/LIGO-M1100296/public}}\BibitemShut {NoStop}%
\bibitem [{\citenamefont {{Aso}}\ \emph {et~al.}(2013)\citenamefont {{Aso}}
  \emph {et~al.}}]{kagra}%
  \BibitemOpen
  \bibfield  {author} {\bibinfo {author} {\bibfnamefont {Y.}~\bibnamefont
  {{Aso}}} \emph {et~al.},\ }\href {\doibase 10.1103/PhysRevD.88.043007}
  {\bibfield  {journal} {\bibinfo  {journal} {\prd}\ }\textbf {\bibinfo
  {volume} {88}},\ \bibinfo {eid} {043007} (\bibinfo {year}
  {2013})}\BibitemShut {NoStop}%
\bibitem [{\citenamefont {Miller}\ \emph {et~al.}(2015)\citenamefont {Miller},
  \citenamefont {Barsotti}, \citenamefont {Vitale}, \citenamefont {Fritschel},
  \citenamefont {Evans},\ and\ \citenamefont {Sigg}}]{Aplus}%
  \BibitemOpen
  \bibfield  {author} {\bibinfo {author} {\bibfnamefont {J.}~\bibnamefont
  {Miller}}, \bibinfo {author} {\bibfnamefont {L.}~\bibnamefont {Barsotti}},
  \bibinfo {author} {\bibfnamefont {S.}~\bibnamefont {Vitale}}, \bibinfo
  {author} {\bibfnamefont {P.}~\bibnamefont {Fritschel}}, \bibinfo {author}
  {\bibfnamefont {M.}~\bibnamefont {Evans}}, \ and\ \bibinfo {author}
  {\bibfnamefont {D.}~\bibnamefont {Sigg}},\ }\href {\doibase
  10.1103/PhysRevD.91.062005} {\bibfield  {journal} {\bibinfo  {journal}
  {\prd}\ }\textbf {\bibinfo {volume} {91}},\ \bibinfo {pages} {062005}
  (\bibinfo {year} {2015})}\BibitemShut {NoStop}%
\bibitem [{\citenamefont {Abbott}\ \emph {et~al.}(2017)\citenamefont {Abbott}
  \emph {et~al.}}]{CosmicExplorer}%
  \BibitemOpen
  \bibfield  {author} {\bibinfo {author} {\bibfnamefont {B.~P.}\ \bibnamefont
  {Abbott}} \emph {et~al.},\ }\href {\doibase 10.1088/1361-6382/aa51f4}
  {\bibfield  {journal} {\bibinfo  {journal} {Classical Quantum Gravity}\
  }\textbf {\bibinfo {volume} {34}},\ \bibinfo {pages} {44001} (\bibinfo {year}
  {2017})}\BibitemShut {NoStop}%
\bibitem [{\citenamefont {Punturo}\ \emph {et~al.}(2010)\citenamefont {Punturo}
  \emph {et~al.}}]{EinsteinTelescope}%
  \BibitemOpen
  \bibfield  {author} {\bibinfo {author} {\bibfnamefont {M.}~\bibnamefont
  {Punturo}} \emph {et~al.},\ }\href
  {http://stacks.iop.org/0264-9381/27/i=19/a=194002} {\bibfield  {journal}
  {\bibinfo  {journal} {Classical Quantum Gravity}\ }\textbf {\bibinfo {volume}
  {27}},\ \bibinfo {pages} {194002} (\bibinfo {year} {2010})}\BibitemShut
  {NoStop}%
\bibitem [{pyg()}]{pygwinc}%
  \BibitemOpen
  \href@noop {} {}\bibinfo {howpublished}
  {\url{https://git.ligo.org/gwinc/pygwinc}}\BibitemShut {NoStop}%
\bibitem [{Note6()}]{Note6}%
  \BibitemOpen
  \bibinfo {note} {This example is found in the {\protect \sc Bilby}\protect
  \xspace {} {\protect \tt git} repository at \protect \url
  {https://git.ligo.org/lscsoft/bilby/blob/master/examples/injection_examples/Australian_detector.py}.}\BibitemShut
  {Stop}%
\bibitem [{\citenamefont {Brown}\ and\ \citenamefont {Freise}(2014)}]{finesse}%
  \BibitemOpen
  \bibfield  {author} {\bibinfo {author} {\bibfnamefont {D.~D.}\ \bibnamefont
  {Brown}}\ and\ \bibinfo {author} {\bibfnamefont {A.}~\bibnamefont {Freise}},\
  }\href {\doibase 10.5281/zenodo.821363} {\enquote {\bibinfo {title}
  {Finesse},}\ } (\bibinfo {year} {2014}),\ \bibinfo {note} {{You can download
  the binaries and source code at
  \url{http://www.gwoptics.org/finesse}.}}\BibitemShut {Stop}%
\bibitem [{\citenamefont {{Logue}}\ \emph {et~al.}(2012)\citenamefont
  {{Logue}}, \citenamefont {{Ott}}, \citenamefont {{Heng}}, \citenamefont
  {{Kalmus}},\ and\ \citenamefont {{Scargill}}}]{logue12}%
  \BibitemOpen
  \bibfield  {author} {\bibinfo {author} {\bibfnamefont {J.}~\bibnamefont
  {{Logue}}}, \bibinfo {author} {\bibfnamefont {C.~D.}\ \bibnamefont {{Ott}}},
  \bibinfo {author} {\bibfnamefont {I.~S.}\ \bibnamefont {{Heng}}}, \bibinfo
  {author} {\bibfnamefont {P.}~\bibnamefont {{Kalmus}}}, \ and\ \bibinfo
  {author} {\bibfnamefont {J.}~\bibnamefont {{Scargill}}},\ }\href {\doibase
  10.1103/PhysRevD.86.044023} {\bibfield  {journal} {\bibinfo  {journal}
  {\prd}\ }\textbf {\bibinfo {volume} {86}},\ \bibinfo {eid} {044023} (\bibinfo
  {year} {2012})}\BibitemShut {NoStop}%
\bibitem [{\citenamefont {{Powell}}\ \emph {et~al.}(2016)\citenamefont
  {{Powell}}, \citenamefont {{Gossan}}, \citenamefont {{Logue}},\ and\
  \citenamefont {{Heng}}}]{powell16}%
  \BibitemOpen
  \bibfield  {author} {\bibinfo {author} {\bibfnamefont {J.}~\bibnamefont
  {{Powell}}}, \bibinfo {author} {\bibfnamefont {S.~E.}\ \bibnamefont
  {{Gossan}}}, \bibinfo {author} {\bibfnamefont {J.}~\bibnamefont {{Logue}}}, \
  and\ \bibinfo {author} {\bibfnamefont {I.~S.}\ \bibnamefont {{Heng}}},\
  }\href {\doibase 10.1103/PhysRevD.94.123012} {\bibfield  {journal} {\bibinfo
  {journal} {\prd}\ }\textbf {\bibinfo {volume} {94}},\ \bibinfo {eid} {123012}
  (\bibinfo {year} {2016})}\BibitemShut {NoStop}%
\bibitem [{\citenamefont {{Powell}}\ \emph {et~al.}(2017)\citenamefont
  {{Powell}}, \citenamefont {{Szczepanczyk}},\ and\ \citenamefont
  {{Heng}}}]{powell17}%
  \BibitemOpen
  \bibfield  {author} {\bibinfo {author} {\bibfnamefont {J.}~\bibnamefont
  {{Powell}}}, \bibinfo {author} {\bibfnamefont {M.}~\bibnamefont
  {{Szczepanczyk}}}, \ and\ \bibinfo {author} {\bibfnamefont {I.~S.}\
  \bibnamefont {{Heng}}},\ }\href {\doibase 10.1103/PhysRevD.96.123013}
  {\bibfield  {journal} {\bibinfo  {journal} {\prd}\ }\textbf {\bibinfo
  {volume} {96}},\ \bibinfo {eid} {123013} (\bibinfo {year}
  {2017})}\BibitemShut {NoStop}%
\bibitem [{\citenamefont {{M{\"u}ller}}\ \emph {et~al.}(2012)\citenamefont
  {{M{\"u}ller}}, \citenamefont {{Janka}},\ and\ \citenamefont
  {{Wongwathanarat}}}]{mueller12}%
  \BibitemOpen
  \bibfield  {author} {\bibinfo {author} {\bibfnamefont {E.}~\bibnamefont
  {{M{\"u}ller}}}, \bibinfo {author} {\bibfnamefont {H.-T.}\ \bibnamefont
  {{Janka}}}, \ and\ \bibinfo {author} {\bibfnamefont {A.}~\bibnamefont
  {{Wongwathanarat}}},\ }\href {\doibase 10.1051/0004-6361/201117611}
  {\bibfield  {journal} {\bibinfo  {journal} {Astron. Astrophys.}\ }\textbf
  {\bibinfo {volume} {537}},\ \bibinfo {eid} {A63} (\bibinfo {year}
  {2012})}\BibitemShut {NoStop}%
\bibitem [{Note7()}]{Note7}%
  \BibitemOpen
  \bibinfo {note} {This example is found in the {\protect \sc Bilby}\protect
  \xspace {} {\protect \tt git} repository at \protect \url
  {https://git.ligo.org/lscsoft/bilby/blob/master/examples/supernova_example/supernova_example.py}.}\BibitemShut
  {Stop}%
\bibitem [{\citenamefont {{Clark}}\ \emph {et~al.}(2014)\citenamefont
  {{Clark}}, \citenamefont {{Bauswein}}, \citenamefont {{Cadonati}},
  \citenamefont {{Janka}}, \citenamefont {{Pankow}},\ and\ \citenamefont
  {{Stergioulas}}}]{clark14}%
  \BibitemOpen
  \bibfield  {author} {\bibinfo {author} {\bibfnamefont {J.}~\bibnamefont
  {{Clark}}}, \bibinfo {author} {\bibfnamefont {A.}~\bibnamefont {{Bauswein}}},
  \bibinfo {author} {\bibfnamefont {L.}~\bibnamefont {{Cadonati}}}, \bibinfo
  {author} {\bibfnamefont {H.-T.}\ \bibnamefont {{Janka}}}, \bibinfo {author}
  {\bibfnamefont {C.}~\bibnamefont {{Pankow}}}, \ and\ \bibinfo {author}
  {\bibfnamefont {N.}~\bibnamefont {{Stergioulas}}},\ }\href {\doibase
  10.1103/PhysRevD.90.062004} {\bibfield  {journal} {\bibinfo  {journal}
  {\prd}\ }\textbf {\bibinfo {volume} {90}},\ \bibinfo {eid} {062004} (\bibinfo
  {year} {2014})}\BibitemShut {NoStop}%
\bibitem [{\citenamefont {{Abbott}}\ \emph {et~al.}(2017)\citenamefont
  {{Abbott}} \emph {et~al.}}]{abbott17_gw170817_postmerger}%
  \BibitemOpen
  \bibfield  {author} {\bibinfo {author} {\bibfnamefont {B.~P.}\ \bibnamefont
  {{Abbott}}} \emph {et~al.},\ }\href {\doibase 10.3847/2041-8213/aa9a35}
  {\bibfield  {journal} {\bibinfo  {journal} {Astrophys. J.}\ }\textbf
  {\bibinfo {volume} {851}},\ \bibinfo {eid} {L16} (\bibinfo {year}
  {2017})}\BibitemShut {NoStop}%
\bibitem [{\citenamefont {{Shibata}}\ and\ \citenamefont
  {{Taniguchi}}(2006)}]{shibata06}%
  \BibitemOpen
  \bibfield  {author} {\bibinfo {author} {\bibfnamefont {M.}~\bibnamefont
  {{Shibata}}}\ and\ \bibinfo {author} {\bibfnamefont {K.}~\bibnamefont
  {{Taniguchi}}},\ }\href {\doibase 10.1103/PhysRevD.73.064027} {\bibfield
  {journal} {\bibinfo  {journal} {\prd}\ }\textbf {\bibinfo {volume} {73}},\
  \bibinfo {eid} {064027} (\bibinfo {year} {2006})}\BibitemShut {NoStop}%
\bibitem [{\citenamefont {{Baiotti}}\ \emph {et~al.}(2008)\citenamefont
  {{Baiotti}}, \citenamefont {{Giacomazzo}},\ and\ \citenamefont
  {{Rezzolla}}}]{baiotti08}%
  \BibitemOpen
  \bibfield  {author} {\bibinfo {author} {\bibfnamefont {L.}~\bibnamefont
  {{Baiotti}}}, \bibinfo {author} {\bibfnamefont {B.}~\bibnamefont
  {{Giacomazzo}}}, \ and\ \bibinfo {author} {\bibfnamefont {L.}~\bibnamefont
  {{Rezzolla}}},\ }\href {\doibase 10.1103/PhysRevD.78.084033} {\bibfield
  {journal} {\bibinfo  {journal} {\prd}\ }\textbf {\bibinfo {volume} {78}},\
  \bibinfo {eid} {084033} (\bibinfo {year} {2008})}\BibitemShut {NoStop}%
\bibitem [{\citenamefont {{Read}}\ \emph {et~al.}(2013)\citenamefont {{Read}}
  \emph {et~al.}}]{read13}%
  \BibitemOpen
  \bibfield  {author} {\bibinfo {author} {\bibfnamefont {J.~S.}\ \bibnamefont
  {{Read}}} \emph {et~al.},\ }\href {\doibase 10.1103/PhysRevD.88.044042}
  {\bibfield  {journal} {\bibinfo  {journal} {\prd}\ }\textbf {\bibinfo
  {volume} {88}},\ \bibinfo {eid} {044042} (\bibinfo {year}
  {2013})}\BibitemShut {NoStop}%
\bibitem [{\citenamefont {{Chatziioannou}}\ \emph {et~al.}(2017)\citenamefont
  {{Chatziioannou}}, \citenamefont {{Clark}}, \citenamefont {{Bauswein}},
  \citenamefont {{Millhouse}}, \citenamefont {{Littenberg}},\ and\
  \citenamefont {{Cornish}}}]{chatziioannou17}%
  \BibitemOpen
  \bibfield  {author} {\bibinfo {author} {\bibfnamefont {K.}~\bibnamefont
  {{Chatziioannou}}}, \bibinfo {author} {\bibfnamefont {J.~A.}\ \bibnamefont
  {{Clark}}}, \bibinfo {author} {\bibfnamefont {A.}~\bibnamefont {{Bauswein}}},
  \bibinfo {author} {\bibfnamefont {M.}~\bibnamefont {{Millhouse}}}, \bibinfo
  {author} {\bibfnamefont {T.~B.}\ \bibnamefont {{Littenberg}}}, \ and\
  \bibinfo {author} {\bibfnamefont {N.}~\bibnamefont {{Cornish}}},\ }\href
  {\doibase 10.1103/PhysRevD.96.124035} {\bibfield  {journal} {\bibinfo
  {journal} {\prd}\ }\textbf {\bibinfo {volume} {96}},\ \bibinfo {eid} {124035}
  (\bibinfo {year} {2017})}\BibitemShut {NoStop}%
\bibitem [{\citenamefont {{Clark}}\ \emph {et~al.}(2016)\citenamefont
  {{Clark}}, \citenamefont {{Bauswein}}, \citenamefont {{Stergioulas}},\ and\
  \citenamefont {{Shoemaker}}}]{clark16}%
  \BibitemOpen
  \bibfield  {author} {\bibinfo {author} {\bibfnamefont {J.~A.}\ \bibnamefont
  {{Clark}}}, \bibinfo {author} {\bibfnamefont {A.}~\bibnamefont {{Bauswein}}},
  \bibinfo {author} {\bibfnamefont {N.}~\bibnamefont {{Stergioulas}}}, \ and\
  \bibinfo {author} {\bibfnamefont {D.}~\bibnamefont {{Shoemaker}}},\ }\href
  {\doibase 10.1088/0264-9381/33/8/085003} {\bibfield  {journal} {\bibinfo
  {journal} {Classical Quantum Gravity}\ }\textbf {\bibinfo {volume} {33}},\
  \bibinfo {eid} {085003} (\bibinfo {year} {2016})}\BibitemShut {NoStop}%
\bibitem [{\citenamefont {{Easter}}\ \emph {et~al.}(2018)\citenamefont
  {{Easter}}, \citenamefont {{Lasky}}, \citenamefont {{Casey}}, \citenamefont
  {{Rezzolla}},\ and\ \citenamefont {{Takami}}}]{easter18}%
  \BibitemOpen
  \bibfield  {author} {\bibinfo {author} {\bibfnamefont {P.~J.}\ \bibnamefont
  {{Easter}}}, \bibinfo {author} {\bibfnamefont {P.~D.}\ \bibnamefont
  {{Lasky}}}, \bibinfo {author} {\bibfnamefont {A.~R.}\ \bibnamefont
  {{Casey}}}, \bibinfo {author} {\bibfnamefont {L.}~\bibnamefont {{Rezzolla}}},
  \ and\ \bibinfo {author} {\bibfnamefont {K.}~\bibnamefont {{Takami}}},\
  }\href@noop {} {\enquote {\bibinfo {title} {{Computing Fast and Reliable
  Gravitational Waveforms of Binary Neutron Star Merger Remnants}},}\ }
  (\bibinfo {year} {2018})\BibitemShut {NoStop}%
\bibitem [{\citenamefont {{Messenger}}\ \emph {et~al.}(2014)\citenamefont
  {{Messenger}}, \citenamefont {{Takami}}, \citenamefont {{Gossan}},
  \citenamefont {{Rezzolla}},\ and\ \citenamefont
  {{Sathyaprakash}}}]{messenger14}%
  \BibitemOpen
  \bibfield  {author} {\bibinfo {author} {\bibfnamefont {C.}~\bibnamefont
  {{Messenger}}}, \bibinfo {author} {\bibfnamefont {K.}~\bibnamefont
  {{Takami}}}, \bibinfo {author} {\bibfnamefont {S.}~\bibnamefont {{Gossan}}},
  \bibinfo {author} {\bibfnamefont {L.}~\bibnamefont {{Rezzolla}}}, \ and\
  \bibinfo {author} {\bibfnamefont {B.~S.}\ \bibnamefont {{Sathyaprakash}}},\
  }\href {\doibase 10.1103/PhysRevX.4.041004} {\bibfield  {journal} {\bibinfo
  {journal} {Phys. Rev. X}\ }\textbf {\bibinfo {volume} {4}},\ \bibinfo {eid}
  {041004} (\bibinfo {year} {2014})}\BibitemShut {NoStop}%
\bibitem [{\citenamefont {{Bose}}\ \emph {et~al.}(2018)\citenamefont {{Bose}},
  \citenamefont {{Chakravarti}}, \citenamefont {{Rezzolla}}, \citenamefont
  {{Sathyaprakash}},\ and\ \citenamefont {{Takami}}}]{bose18}%
  \BibitemOpen
  \bibfield  {author} {\bibinfo {author} {\bibfnamefont {S.}~\bibnamefont
  {{Bose}}}, \bibinfo {author} {\bibfnamefont {K.}~\bibnamefont
  {{Chakravarti}}}, \bibinfo {author} {\bibfnamefont {L.}~\bibnamefont
  {{Rezzolla}}}, \bibinfo {author} {\bibfnamefont {B.~S.}\ \bibnamefont
  {{Sathyaprakash}}}, \ and\ \bibinfo {author} {\bibfnamefont {K.}~\bibnamefont
  {{Takami}}},\ }\href {\doibase 10.1103/PhysRevLett.120.031102} {\bibfield
  {journal} {\bibinfo  {journal} {\prl}\ }\textbf {\bibinfo {volume} {120}},\
  \bibinfo {eid} {031102} (\bibinfo {year} {2018})}\BibitemShut {NoStop}%
\bibitem [{\citenamefont {{Takami}}\ \emph {et~al.}(2015)\citenamefont
  {{Takami}}, \citenamefont {{Rezzolla}},\ and\ \citenamefont
  {{Baiotti}}}]{takami15}%
  \BibitemOpen
  \bibfield  {author} {\bibinfo {author} {\bibfnamefont {K.}~\bibnamefont
  {{Takami}}}, \bibinfo {author} {\bibfnamefont {L.}~\bibnamefont
  {{Rezzolla}}}, \ and\ \bibinfo {author} {\bibfnamefont {L.}~\bibnamefont
  {{Baiotti}}},\ }\href {\doibase 10.1103/PhysRevD.91.064001} {\bibfield
  {journal} {\bibinfo  {journal} {\prd}\ }\textbf {\bibinfo {volume} {91}},\
  \bibinfo {eid} {064001} (\bibinfo {year} {2015})}\BibitemShut {NoStop}%
\bibitem [{Note8()}]{Note8}%
  \BibitemOpen
  \bibinfo {note} {This example is found in the {\protect \sc Bilby}\protect
  \xspace {} {\protect \tt git} repository at \protect \url
  {https://git.ligo.org/lscsoft/bilby/blob/master/examples/other_examples/hyper_parameter_example.py}.}\BibitemShut
  {Stop}%
\bibitem [{\citenamefont {{Heger}}\ \emph {et~al.}(2003)\citenamefont
  {{Heger}}, \citenamefont {{Fryer}}, \citenamefont {{Woosley}}, \citenamefont
  {{Langer}},\ and\ \citenamefont {{Hartmann}}}]{Heger2003}%
  \BibitemOpen
  \bibfield  {author} {\bibinfo {author} {\bibfnamefont {A.}~\bibnamefont
  {{Heger}}}, \bibinfo {author} {\bibfnamefont {C.~L.}\ \bibnamefont
  {{Fryer}}}, \bibinfo {author} {\bibfnamefont {S.~E.}\ \bibnamefont
  {{Woosley}}}, \bibinfo {author} {\bibfnamefont {N.}~\bibnamefont {{Langer}}},
  \ and\ \bibinfo {author} {\bibfnamefont {D.~H.}\ \bibnamefont {{Hartmann}}},\
  }\href {\doibase 10.1086/375341} {\bibfield  {journal} {\bibinfo  {journal}
  {Astrophys. J.}\ }\textbf {\bibinfo {volume} {591}},\ \bibinfo {pages} {288}
  (\bibinfo {year} {2003})}\BibitemShut {NoStop}%
\bibitem [{\citenamefont {{Woosley}}\ and\ \citenamefont
  {{Heger}}(2015)}]{Woosley2015}%
  \BibitemOpen
  \bibfield  {author} {\bibinfo {author} {\bibfnamefont {S.~E.}\ \bibnamefont
  {{Woosley}}}\ and\ \bibinfo {author} {\bibfnamefont {A.}~\bibnamefont
  {{Heger}}},\ }in\ \href {\doibase 10.1007/978-3-319-09596-7_7} {\emph
  {\bibinfo {booktitle} {Very Massive Stars in the Local Universe}}},\ \bibinfo
  {series} {Astrophysics and Space Science Library}, Vol.\ \bibinfo {volume}
  {412},\ \bibinfo {editor} {edited by\ \bibinfo {editor} {\bibfnamefont
  {J.~S.}\ \bibnamefont {{Vink}}}}\ (\bibinfo {year} {2015})\ p.\ \bibinfo
  {pages} {199}\BibitemShut {NoStop}%
\bibitem [{\citenamefont {Jones}\ \emph {et~al.}(2001)\citenamefont {Jones}
  \emph {et~al.}}]{scipy}%
  \BibitemOpen
  \bibfield  {author} {\bibinfo {author} {\bibfnamefont {E.}~\bibnamefont
  {Jones}} \emph {et~al.},\ }\href {http://www.scipy.org/} {\enquote {\bibinfo
  {title} {{SciPy}: Open source scientific tools for {Python}},}\ } (\bibinfo
  {year} {2001})\BibitemShut {NoStop}%
\bibitem [{\citenamefont {Travis~E}()}]{numpy}%
  \BibitemOpen
  \bibfield  {author} {\bibinfo {author} {\bibfnamefont {O.}~\bibnamefont
  {Travis~E}},\ }\href@noop {} {\enquote {\bibinfo {title} {{}a guide to
  numpy},}\ }\BibitemShut {NoStop}%
\bibitem [{\citenamefont {McKinney}(2010)}]{pandas}%
  \BibitemOpen
  \bibfield  {author} {\bibinfo {author} {\bibfnamefont {W.}~\bibnamefont
  {McKinney}},\ }\href@noop {} {\bibfield  {journal} {\bibinfo  {journal}
  {Proceedings of the 9th Python in Science Conference}\ }\textbf {\bibinfo
  {volume} {51-56}} (\bibinfo {year} {2010})}\BibitemShut {NoStop}%
\bibitem [{\citenamefont {Hunter}(2007)}]{matplotlib}%
  \BibitemOpen
  \bibfield  {author} {\bibinfo {author} {\bibfnamefont {J.~D.}\ \bibnamefont
  {Hunter}},\ }\href {\doibase 10.1109/MCSE.2007.55} {\bibfield  {journal}
  {\bibinfo  {journal} {Computing In Science \& Engineering}\ }\textbf
  {\bibinfo {volume} {9}},\ \bibinfo {pages} {90} (\bibinfo {year}
  {2007})}\BibitemShut {NoStop}%
\bibitem [{\citenamefont {Foreman-Mackey}(2016)}]{corner}%
  \BibitemOpen
  \bibfield  {author} {\bibinfo {author} {\bibfnamefont {D.}~\bibnamefont
  {Foreman-Mackey}},\ }\href {\doibase 10.21105/joss.00024} {\bibfield
  {journal} {\bibinfo  {journal} {The Journal of Open Source Software}\
  }\textbf {\bibinfo {volume} {24}} (\bibinfo {year} {2016}),\
  10.21105/joss.00024}\BibitemShut {NoStop}%
\bibitem [{\citenamefont {{G{\'o}rski}}\ \emph {et~al.}(2005)\citenamefont
  {{G{\'o}rski}}, \citenamefont {{Hivon}}, \citenamefont {{Banday}},
  \citenamefont {{Wandelt}}, \citenamefont {{Hansen}}, \citenamefont
  {{Reinecke}},\ and\ \citenamefont {{Bartelmann}}}]{healpy}%
  \BibitemOpen
  \bibfield  {author} {\bibinfo {author} {\bibfnamefont {K.~M.}\ \bibnamefont
  {{G{\'o}rski}}}, \bibinfo {author} {\bibfnamefont {E.}~\bibnamefont
  {{Hivon}}}, \bibinfo {author} {\bibfnamefont {A.~J.}\ \bibnamefont
  {{Banday}}}, \bibinfo {author} {\bibfnamefont {B.~D.}\ \bibnamefont
  {{Wandelt}}}, \bibinfo {author} {\bibfnamefont {F.~K.}\ \bibnamefont
  {{Hansen}}}, \bibinfo {author} {\bibfnamefont {M.}~\bibnamefont
  {{Reinecke}}}, \ and\ \bibinfo {author} {\bibfnamefont {M.}~\bibnamefont
  {{Bartelmann}}},\ }\href {\doibase 10.1086/427976} {\bibfield  {journal}
  {\bibinfo  {journal} {Astrophys. J.}\ }\textbf {\bibinfo {volume} {622}},\
  \bibinfo {pages} {759} (\bibinfo {year} {2005})}\BibitemShut {NoStop}%
\bibitem [{dee()}]{deepdish}%
  \BibitemOpen
  \href@noop {} {}\bibinfo {howpublished}
  {\url{https://github.com/uchicago-cs/deepdish}}\BibitemShut {NoStop}%
\bibitem [{\citenamefont {{Robitaille}}\ \emph {et~al.}(2013)\citenamefont
  {{Robitaille}} \emph {et~al.}}]{astropy1}%
  \BibitemOpen
  \bibfield  {author} {\bibinfo {author} {\bibfnamefont {T.~P.}\ \bibnamefont
  {{Robitaille}}} \emph {et~al.},\ }\href {\doibase
  10.1051/0004-6361/201322068} {\bibfield  {journal} {\bibinfo  {journal}
  {Astron. Astrophys.}\ }\textbf {\bibinfo {volume} {558}},\ \bibinfo {eid}
  {A33} (\bibinfo {year} {2013})},\ \Eprint {http://arxiv.org/abs/1307.6212}
  {arXiv:1307.6212 [astro-ph.IM]} \BibitemShut {NoStop}%
\bibitem [{\citenamefont {{Price-Whelan}}\ \emph {et~al.}(2018)\citenamefont
  {{Price-Whelan}} \emph {et~al.}}]{astropy2}%
  \BibitemOpen
  \bibfield  {author} {\bibinfo {author} {\bibfnamefont {A.~M.}\ \bibnamefont
  {{Price-Whelan}}} \emph {et~al.},\ }\href {\doibase 10.3847/1538-3881/aabc4f}
  {\bibfield  {journal} {\bibinfo  {journal} {Astron. J.}\ }\textbf {\bibinfo
  {volume} {156}},\ \bibinfo {eid} {123} (\bibinfo {year} {2018})}\BibitemShut
  {NoStop}%
\end{thebibliography}%

\end{document}